\documentclass{JHEP3}
\usepackage[english]{babel}
\usepackage{cite}

\usepackage{gensymb}

\usepackage{epsf}
\usepackage{amssymb}
\usepackage{amsmath}
\usepackage{amsfonts}
\usepackage{psfrag,epsfig,graphicx,graphics}

\def\Tr{\rm Tr}

\title{Towards a complete next-to-logarithmic description of forward exclusive diffractive dijet
electroproduction at HERA: real corrections}

\author{R.~Boussarie
\\
Physics Department, Brookhaven National Laboratory, Upton, NY 11973, USA\\
Email: \email{rboussarie@bnl.gov}}
\author{A.~V.~Grabovsky
\\
Budker Institute of Nuclear Physics, 11, Lavrenteva avenue, 630090, Novosibirsk, Russia {\em \&} \\
Novosibirsk State University, 630090, 2, Pirogova street, Novosibirsk, Russia\\
Email: \email{A.V.Grabovsky@inp.nsk.su}}
\author{L.~Szymanowski
\\
National Centre for Nuclear Research (NCBJ), 02-093 Warsaw, Poland\\
Email: \email{Lech.Szymanowski@ncbj.gov.pl}}
\author{S. Wallon
\\
Laboratoire de Physique Th\'eorique (UMR 8627), CNRS, Univ. Paris-Sud, Universit\'e Paris-Saclay, 91405, Orsay, France {\em \&} \\
Sorbonne Universit\'e, Facult\'e de Physique, 4 place Jussieu, 75252 Paris Cedex 05, France\\
Email: \email{wallon@th.u-psud.fr}}

\abstract{We studied the $ep\rightarrow ep+2jets$ diffractive cross section with ZEUS phase
space.
Neglecting the
$t$-channel momentum in the Born and gluon dipole impact factors, we
calculated the corresponding contributions to the cross section
differential in $\beta=\frac{Q^{2}}{Q^{2}+M_{2jets}^{2}}$ and the angle $\phi$
between the leptonic and hadronic planes. The gluon dipole contribution was
obtained in the exclusive $k_{t}$-algorithm with the exclusive cut $y_{cut}=0.15$ in the small $y_{cut}$
approximation. In the collinear approximation we canceled singularities
between real and virtual contributions to the $q\bar{q}$ dipole configuration,
keeping the exact $y_{cut}$ dependency. We used the Golec-Biernat -
W\"usthoff (GBW) 
parametrization for the dipole
matrix element and linearized the double dipole contributions. The results
give roughly $\frac{1}{2}$ of the observed cross section for small $\beta$ and
coincides with it for large $\beta.$}

\begin{document}

\pagestyle{empty}
\newpage

\mbox{}

\pagestyle{plain}

\setcounter{page}{1}

\section{Introduction}

One of the main outcomes of the HERA research program is the evidence and detailed study of diffractive processes. Indeed,  almost 10~\%  of the $\gamma^* p \to hadrons$ deep inelastic scattering (DIS) events were shown to contain
 a rapidity gap in the detectors between the proton remnants $Y$
and the hadrons $X$
coming from the fragmentation region of the initial virtual photon, namely the process was shown to look like $\gamma^* p \to X \, Y$. These
 diffractive deep inelastic scattering (DDIS) events were revealed and extensively studied by H1 and ZEUS collaborations~\cite{Aktas:2006hx,Aktas:2006hy,Chekanov:2004hy,Chekanov:2005vv,Aaron:2010aa,Aaron:2012ad,Chekanov:2008fh,Aaron:2012hua}.
The existence of a rapidity gap between the diffractive state $X$ and the proton remnants, with vacuum quantum numbers in $t-$channel, is a natural place for a Pomeron-like description. Two types of approaches have been developed. 

First, based on 
 the existence of a hard scale (the photon virtuality $Q^2$ for DIS),  a collinear QCD factorization theorem was derived~\cite{Collins:1997sr} and applied successfully to diffractive processes. 
 For inclusive diffraction, this theorem is usually applied with so-called resolved Pomeron models, where one introduces distributions of partons inside the Pomeron, similarly to the usual parton distribution functions for proton in DIS, convoluted with hard matrix elements. 
 In the framework of collinear factorization diffractive dijet photoproduction was calculated in \cite{Guzey:2016tek} and
\cite{Guzey:2016awf} in NLO pQCD, where the authors observed collinear factorization breaking. To describe the data it was
necessary to introduce a model for the suppression factor or gap survival probability. They demonstrated that a global
suppression factor or a model depending on the light cone momentum fraction and the flavour of the interacting parton
describe the HERA data. Inclusive dijet photoproduction was also studied in this framework and was shown to be very
sensitive to the details of nuclear PDFs in the Pb-Pb ultraperipheral collisions in the LHC kinematics 
\cite{Guzey:2018dlm},  \cite{Guzey:2019kik}.

 Second, it is natural at very high energies to view the process as the coupling of a Pomeron with the diffractive state $X$ of invariant mass $M$. In the rest of this paper, we generically call such descriptions as high-energy factorization pictures. In DDIS case, 
for low values of $M^2$, $X$ can be modeled by a  $q \bar{q}$ pair, while for larger values of $M^2,$
the cross section with an additional produced gluon, i.e.\ $X=q \bar{q} g,$
is enhanced. 
 A good description of HERA data for diffraction was achieved in such a model~\cite{Bartels:1998ea}, in which the Pomeron was 
 described by a two-gluon exchange. 

In the present paper, 
we study in detail the cross section for exclusive dijet  electroproduction in diffraction, as was recently reported by ZEUS~\cite{Abramowicz:2015vnu}. A first theoretical study of such processes within a  high-energy factorization picture was performed in \cite{Bartels:1996tc}, in a leading order (LO) approximation in which the dijet was made of a $q \bar{q}$ pair. 

Our aim is to make a description of the same process, relying now on our complete next-to-leading order (NLO) description of the direct coupling of the Pomeron
to the diffractive $X$ state, obtained in refs.~\cite{Boussarie:2014lxa,Boussarie:2016ogo}, and further extended to the case of a light vector meson in ref.~\cite{Boussarie:2016bkq}. 
In our approach, the Pomeron is understood as
a color singlet QCD shockwave, in the spirit of Balitsky's high energy operator expansion~\cite{Balitsky:1995ub, Balitsky:1998kc, Balitsky:1998ya, Balitsky:2001re}
or in its color glass condensate formulation~\cite{JalilianMarian:1997jx,JalilianMarian:1997gr,JalilianMarian:1997dw,JalilianMarian:1998cb,Kovner:2000pt,Weigert:2000gi,Iancu:2000hn,Iancu:2001ad,Ferreiro:2001qy}.

The exclusive diffractive production of a dijet will be a key process for the physics at the Electron-Ion Collider (EIC) at small $x$. Indeed, it was proven to probe the dipole Wigner distribution~\cite{Hatta:2016dxp}. Several recent studies have been performed in order to build precise target matrix elements for EIC phenomenology~\cite{Altinoluk:2015dpi, Mantysaari:2019csc, Salazar:2019ncp} and for Ultraperipheral collisions at the LHC~\cite{Hagiwara:2017fye}. The gluon Wigner distributions probed by our process can describe a cold nuclear origin for elliptic anisotropies, as studied for dilute-dense collisions~\cite{Iancu:2017fzn, Hagiwara:2017ofm}. Finally the (subeikonal) target spin asymmetry for dijet production was proven to give a direct access to the gluon orbital angular momentum in the target~\cite{Ji:2016jgn,Hatta:2016aoc}. In this paper, we are interested in building accurate descriptions of the final state via a jet algorithm, to be combined later with the target matrix elements in the aforementioned studies for future precise EIC predictions

We present explicit formulas for Born $ep\rightarrow ep^{\prime}+2jets$ cross
section allowed by HERA kinematics. We argue that for Born production
mechanism, HERA selection cuts for diffractive DIS~\cite{Abramowicz:2015vnu}
severely reduce contributions from jets in the aligned configuration since simultaneous
restrictions on $p_{\bot jet}>2~{\rm GeV}$ and $M_{2jets}>5~{\rm GeV}$ forbid a jet with a
very small longitudinal momentum fraction of the photon. As is known~\cite{GolecBiernat:2001hab}, the aligned jets give the dominant contribution to the
cross section, which in the presently studied kinematics is cut off. Thanks to these cuts the typical transverse
energy scale in the Born jet impact factor is greater than the $t$-channel
transverse momentum scale set by the saturation scale $Q_{s}$ determined by
the proton matrix element. As a result, we can expand in the $t$-channel
transverse momentum in the impact factor and analytically take integrals for
the $\gamma p$ cross section. This naturally gives the leading power $\sim
Q_{s}^{4}\sim W^{2\lambda}$ behavior of the cross section (where $1+\lambda$ is the pomeron intercept) unlike $\sim
Q_{s}^{2}\sim W^{\lambda}$ for the aligned jets~\cite{GolecBiernat:2001hab}
describing large dipoles and saturation. We called this procedure
\textquotedblleft small $Q_{s}$\textquotedblright\ or $``$%
BFKL-like\textquotedblright\ approximation.\ 

Next, we study the real radiative corrections. According to the exclusive $k_{t}$ jet
algorithm~\cite{Brown:1991hx,Catani:1991hj} used in the ZEUS data analysis~\cite{Abramowicz:2015vnu}, these corrections come from the $\sim\sqrt{y_{cut}}$-wide border of the
Dalitz plot (see figure \ref{dalitzqqbar}), with $y_{cut}=0.15$ being the
algorithm parameter. One can symmetrically divide this area into 3 subareas
with predominantly $q-(\bar{q}g),$ $\bar{q}-(qg),$ and $g-(q\bar{q})$ jets,
where one of the jets is made of $\bar{q}g,$ $qg,$ or $q\bar{q}$
correspondingly. At large $M_{2jets}$ the third region gives enhanced
contribution since in such kinematics a subdiagram with a $t$-channel gluon
has large $s=M_{2jets}^{2}.$

Most of the\ real production matrix elements were calculated in
ref.~\cite{Boussarie:2016ogo} in arbitrary kinematics. We have obtained here the
remaining ones and we present them in appendix~\ref{append:interference}. The real production matrix
elements have soft and collinear divergencies in the first two regions while
the contribution of the third region is finite. Integrating the singular parts
over the first two regions, we cancel the singularities with the singular
contribution of the virtual part from ref.~\cite{Boussarie:2016ogo}. As a result,
we have the contribution of soft and collinear gluons to the 2 jet cross
section in the $k_{t}$ algorithm. Since the divergent contributions factorize
as the Born cross section times the collinear singular factor, the validity
criteria of the small $Q_{s}$ approximation for such contributions are the
same as for the Born cross section. Therefore we used this approximation to
take the inner integrals in the $\gamma p$ cross section. The average value of
this correction is about 10\%. However, we noticed that the small $y_{cut}$
expansion of this contribution is very inaccurate since $\ln^{2}y_{cut},$ $\ln
y_{cut},$ and the constant contributions together are of the order of the next term
$\sim\sqrt{y_{cut}}=0.39,$ which is the true expansion parameter. Although we
calculated this contribution exactly in $y_{cut},$ all other (nonsingular)
contributions are $\sim\sqrt{y_{cut}}$ geometrically. Therefore this term
alone can not be a good approximation. Instead one can look at it as at a
subtraction term for future full numerical calculation.

Among the nonsingular contributions there are ones with the gluon emitted
before the shockwave. Suppose for definiteness that it was emitted from the
quark and we consider the second, $\bar{q}-(qg)$ region. In such a
contribution the invariant mass of the $qg$ pair is small $\sim\sqrt{y_{cut}},$
and the only hard scale in the quark propagator between the photon and the
gluon vertices comes from the $t$-channel. It means that one cannot neglect the
$t$-channel momentum in the impact factor, i.e. the small $Q_{s}$
approximation is inapplicable. In other words this correction is very
sensitive to $Q_{s}$. In general, one can say that if one experimentally
restricts from below both the mass of the dijet system and the transverse
momentum of the jet so that the aligned jets are cut off from the Born cross
section, the radiative corrections will greatly depend on saturation effects.

For a generic, roughly symmetric, dijet configuration in the third region with roughly $\frac{1}{2}$ of
the photon's longitudinal momentum taken by the gluon and roughly $\frac{1}%
{4}$ taken by the quark and antiquark each, the typical transverse energy
scale in the impact factor is determined by the same parameters as in the Born
one: the photon virtuality $Q$, $M_{2jets},$ and the experimental cut $p_{\bot
jet\min}.$ Therefore one can also try calculating the contribution of such
gluon dipoles in the small $Q_{s}$ approximation. The validity of this
approximation for the configuration when the $(q\bar{q})$ jet itself has the
aligned structure is not justified, however, since then the quark or
antiquark's part of the longitudinal momentum of the pair becomes a new small
parameter. Such a situation happens in the corners of the third $g-(q\bar{q})$
area in the Dalitz plot since in these corners the invariant mass of $qg$ or
$\bar{q}g$ becomes small and we return to the situation discussed in the
previous paragraph.

Anyway in this paper we have calculated the contribution of all real
radiative corrections from the third $g-(q\bar{q})$ area in the Dalitz plot,
i.e. the gluon dipole contribution in the small $Q_{s}$ approximation, i.e.
expanding the impact factors in the $t$-channel momenta. The error of our
result comes from the corners of the phase space discussed above and its
numerical value will be judged from comparison of our result to the future
full numerical calculation. This difference will be related to saturation effects.

This paper is organized as follows. The second part discusses kinematics and yields the LO
computation of the cross section, including its leptonic part in section~\ref{SubSec:leptonic-part},
hadronic part in section~\ref{SubSec:hadronic-part}, HERA acceptance in section~\ref{SubSec:Experimental-cut}, small $Q_{s}$
approximation in section~\ref{SubSec:BFKL-Approximation} and analysis of the result in section~\ref{SubSec:LO-analysis}. The
third part discusses the NLO real corrections including the $k_{t}$-jet
algorithm in section~\ref{SubSec:jet-algorithm-3p}, $q-(\bar{q}g)$ and $\bar{q}-(qg)$ dipoles in section~\ref{SubSec:qg-and-qbarg-jets} and $g-(q\bar{q})$ dipole in section~\ref{SubSec:qqbar-jets}. The conclusion summarizes the paper.
Appendix~\ref{append:Scaling} contains discussion of aligned vs symmetric jet contributions to the
Born cross section. Appendix~\ref{append:interference} presents the dipole - double dipole
interference impact factors for real correction.
 Appendix~\ref{append:normalization} discusses the overall normalization and matching to non-perturbative distributions in the Golec-Biernat W\"usthoff formulation of DDIS.

\section{Kinematics and LO results}
\label{Sec:kin-LO}

\subsection{Leptonic part}
\label{SubSec:leptonic-part}

We will use hereafter  the light cone vectors $n_{1}$ and $n_{2}$, defined as
\begin{equation}
n_{1} \equiv \left(  1,0_{\bot},1\right)  ,\quad n_{2} \equiv \frac{1}{2}\left(  1,0_{\bot
},-1\right)  ,\quad n_{1}^{+}=n_{2}^{-}=(n_{1} \cdot n_{2})=1 .
\end{equation}
For any vector $p$ we note
\begin{align}
p^{+}  & = p_{-} \equiv (p \cdot n_{2})=\frac{1}{2}\left(  p^{0}+p^{3}\right)  ,\qquad p_{+}=p^{-} \equiv (p \cdot n_{1})=p^{0}-p^{3},
\\
p & =p^{+}n_{1}+p^{-}n_{2}+p_{\bot},
\end{align}
so that
\begin{equation}
(p\,\cdot k)=p^{\mu}k_{\mu}=p^{+}k^{-}+p^{-}k^{+}+(p_\bot \cdot k_\bot)=p_{+}k_{-}+p_{-}k_{+}-(\vec{p} \cdot \vec{k}).
\end{equation}

The DIS kinematic variables read%
\begin{align}
s  &  =(p_{0}+k)^{2},\quad q=k-k^{\prime},\quad Q^{2}=-q^{2},\quad
y=\frac{(p_{0}q)}{(p_{0}k)}=\frac{W^{2}+Q^{2}}{s},\quad\bar{y}=1-y,\\
W^{2}  &  =(p_{0}+q)^{2}=2p_{0}q-Q^{2},\quad\frac{dk^{\prime+}d^{2}k_{\bot
}^{\prime}}{k^{^{\prime}+}}=2\pi\frac{dW^{2}dQ^{2}}{2s},
\end{align}
where $p_{0},k,k^{\prime}$ and $q$ are the proton, initial electron, final
electron and photon's momenta and we integrated out the azimuthal angle of the
scattered electron w.r.t. the initial electron via overall rotational
invariance. The cross section for diffractive dijet production reads%
\begin{equation}
d\sigma_{ep}=\frac{4\pi\alpha}{Q^{4}}J_{\mu\nu}\frac{dk^{\prime+}d^{2}k_{\bot
}^{\prime}}{(2\pi)^{3}2k^{^{\prime}+}}\frac{I_{\gamma p}}{I_{ep}}%
\,d\sigma_{\gamma^{\ast}p}^{\mu\nu}\,.
\end{equation}
Here%
\begin{equation}
d\sigma_{\gamma^{\ast}p}^{\mu\nu}=M^{\mu}M^{\nu\ast}\frac{d\rho_{p,jets}%
}{4I_{\gamma p}}%
\end{equation}
is the $\gamma^\ast$-proton cross section, obtained from the $\gamma^\ast$-proton
scattering amplitude $M^{\mu},$ and
\begin{equation}
\frac{I_{\gamma p}}{I_{ep}}=y,\quad J_{\mu\nu}=\frac{1}{2}tr(\hat{k}%
\gamma_{\mu}\hat{k}^{\prime}\gamma_{\nu})=2(k_{\mu}k_{\nu}^{\prime}+k_{\mu
}^{\prime}k_{\nu}-(kk^{\prime})g_{\mu\nu}).
\end{equation}
The photon polarization vectors read
\begin{equation}
\label{polarisation}
e^{(0)}   =\frac{Q}{(qp_{0})}(p_{0}+\frac{(qp_{0})}{Q^{2}}q),\quad
e^{(y)}\equiv e^{(-1)}=\frac{n}{\sqrt{-n^{2}}}, \quad
e^{(x)}    \equiv e^{(1)}=\frac{\tilde{p}_{q}}{\sqrt{-\tilde{p}_{q}^{2}}%
}\,,
\end{equation}
where
\begin{align}
n_{\mu} &=\varepsilon
_{\mu\nu\alpha\beta}p_{q}^{\nu}q^{\alpha}p_{0}^{\beta},\quad {\rm and} \quad \tilde{p}_{q}\equiv p_{q\perp}=p_{q}-q\frac{(p_{0}p_{q})}{(p_{0}q)}-p_{0}\left(
\frac{(qp_{q})}{(p_{0}q)}+\frac{Q^{2}(p_{0}p_{q})}{(p_{0}q)^{2}}\right)  \,.
\end{align}
These polarization vectors obey the identity%
\begin{equation}
e_{\mu}^{x}e_{\nu}^{x\ast}+e_{\mu}^{y}e_{\nu}^{y\ast}=e_{\mu}^{0}e_{\nu
}^{0\ast}-g_{\mu\nu}+\frac{q_{\mu}q_{\nu}}{q^{2}}.
\end{equation}
Hereafter, we label the polarizations using Latin indices, while greek letters are used for Lorentz indices. We get%
\begin{equation}
\label{prop-Jab}
J^{ab}   =(-1)^{a+b}J^{\mu\nu}e_{\mu}^{(a)\ast}e_{\nu}^{(b)}=4(-1)^{a+b}%
(ke^{(a)\ast})(ke^{(b)})-(-1)^{a}Q^{2}\delta^{ab}\,.
\end{equation}
Denoting
\begin{equation}
 d\sigma^{ab}    =d\sigma_{\gamma^{\ast}p}^{\mu\nu}e_{\mu}^{(a)}e_{\nu
}^{(b)\ast}\,,
\end{equation}
we thus have 
\begin{equation}
 J_{\mu \nu} d\sigma_{\gamma^{\ast}p}^{\mu\nu} = J^{ab} d\sigma^{ab} \,.
\end{equation}

In our light-cone frame
\begin{align}
p_{0}  &  =p_{0}^{-}n_{2},\quad q=p_{\gamma}^{+}n_{1}-\frac{Q^{2}}{2p_{\gamma
}^{+}}n_{2},\quad W^{2}+Q^{2}=2p_{0}^{-}p_{\gamma}^{+}%
,\label{proton-photon-frame1}\\
k  &  =p_{e}^{+}n_{1}+\frac{\vec{k}^{\,\,2}}{2p_{e}^{+}}n_{2}+k_{\bot},\quad
k_{\bot}^{i}=\left\vert \vec{k}\right\vert \left(
\begin{array}
[c]{c}%
\cos\phi\\
\sin\phi
\end{array}
\right)  ,\quad s=2p_{e}^{+}p_{0}^{-},\quad\vec{k}^{\,\,2}=\frac{Q^{2}}{y^{2}%
}\bar{y},\\
p_{q}  &  =xp_{\gamma}^{+}n_{1}+\frac{\vec{p}_{q}^{\,\,2}}{2xp_{\gamma}^{+}%
}n_{2}+p_{q\bot},\quad\tilde{p}_{q}^{2}=-\vec{p}_{q}^{\,\,2},\quad n^{2}%
=-\vec{p}_{q}^{\,\,2}\frac{(ys)^{2}}{4}. \label{proton-photon-frame-2}%
\end{align}
It is the frame where the photon and the proton are back-to-back, and the $z$ axis is
along the direction of the photon momentum. The photoproduction cross section \cite{Boussarie:2016ogo}
was calculated in this frame. Hence%
\begin{align}
J^{xx}  &  =\frac{2Q^{2}}{y^{2}}\left[  \bar{y}+\frac{y^{2}}{2}+\bar{y}%
\cos(2\phi)\right]  ,\quad J^{00}=\frac{4Q^{2}}{y^{2}}\bar{y}\,,\\
J^{yy}  &  =\frac{2Q^{2}}{y^{2}}\left[  \bar{y}+\frac{y^{2}}{2}-\bar{y}%
\cos(2\phi)\right]  ,\quad J^{0x}=J^{x0}=\frac{2Q^{2}}{y^{2}}\left(
2-y\right)  \sqrt{\bar{y}}\cos\phi \,,
\end{align}
and%
\begin{equation}
\label{sigmagammaPtoeP}
d\sigma_{ep}=\frac{\alpha y}{4\pi}\frac{dW^{2}dQ^{2}}{sQ^{4}}\,\,J^{ab}%
d\sigma_{\gamma^{\ast}p}^{ab},\quad d\sigma_{\gamma^{\ast}p}^{ab}%
=\frac{d\sigma_{\gamma^{\ast}p}^{ab}}{dxd\vec{p}_{q}d\vec{p}_{\bar{q}}}%
dxd\vec{p}_{q}d\vec{p}_{\bar{q}}\,. %
\end{equation}

\subsection{Hadronic part}
\label{SubSec:hadronic-part}

The density matrix for the cross section in our frame was obtained
in (5.21-23) of ref.~\cite{Boussarie:2016ogo}. To get the proper normalization we have to
multiply all cross sections in ref.~\cite{Boussarie:2016ogo} by $\frac{1}%
{2(2\pi)^{4}}$ as is discussed in appendix \ref{append:normalization}.
The LO cross sections in our frame read \qquad\
\begin{align}
\hspace{-0.2cm}\left.  \frac{d\sigma_{0TT}^{ij}}{dxd\vec{p}_{q}d\vec{p}%
_{\bar{q}}}\right\vert _{t=0}  &  =\frac{1}{2(2\pi)^{4}}\frac{\alpha Q_{q}%
^{2}}{(2\pi)^{4}N_{c}}[((1-2x)^{2}-1)e^{(x)i}e^{(x)j}-g_{\bot}^{ij}]\nonumber \\
& \times \frac
{1}{\vec{p}_{q\bar{q}}^{\,\,2}}\left\vert \int\frac{d^{2}p_{\bot}(\vec
{p}_{q\bar{q}}\vec{p})}{\vec{p}^{\,\,2}+x\bar{x}Q^{2}}\mathbf{F}(p_{\bot
}+\frac{p_{q\bar{q}\bot}}{2})\right\vert ^{2},\label{sigmatt0}
\\
\left.  \frac{d\sigma_{0LL}}{dxd\vec{p}_{q}d\vec{p}_{\bar{q}}}\right\vert
_{t=0}  &  =\frac{1}{2(2\pi)^{4}}\frac{4\alpha Q_{q}^{2}}{(2\pi)^{4}N_{c}%
}x^{2}\bar{x}^{2}Q^{2}\left\vert \int\frac{d^{2}p_{\bot}}{\vec{p}%
^{\,\,2}+x\bar{x}Q^{2}}\mathbf{F}(p_{\bot}+\frac{p_{q\bar{q}}{}_{\bot}}%
{2})\right\vert ^{2},\label{sigmall}\\
\left.  \frac{d\sigma_{0TL}^{i}e_{i\bot}^{(x)}}{dxd\vec{p}_{q}d\vec{p}%
_{\bar{q}}}\right\vert _{t=0}  &  =\frac{1}{2(2\pi)^{4}}\frac{2\alpha
_{\mathrm{em}}Q_{q}^{2}}{(2\pi)^{4}N_{c}}\frac{x\bar{x}(\bar{x}-x)Q}%
{\sqrt{\vec{p}_{q\bar{q}}^{\,\,2}}}\int\frac{d^{2}\vec{p}_{1}\mathbf{F}%
(p_{1\bot}+\frac{p_{q\bar{q}\bot}}{2})}{\vec{p}_{1}^{\,\,2}+x\bar{x}Q^{2}%
} \nonumber \\
& \times \left[  \int\frac{d^{2}\vec{p}(\vec{p}_{q\bar{q}}\vec{p})}{\vec{p}%
^{\,\,2}+x\bar{x}Q^{2}}\mathbf{F}(p_{\bot}+\frac{p_{q\bar{q}\bot}}{2})\right]
^{\ast},
\end{align}
and the total transverse cross section reads%
\begin{equation}
\hspace{-0.2cm}\left.  \frac{d\sigma_{0TT}}{dxd\vec{p}_{q}d\vec{p}_{\bar{q}}%
}\right\vert _{t=0}=\frac{1}{2}\frac{1}{2(2\pi)^{4}}\frac{\alpha Q_{q}^{2}%
}{(2\pi)^{4}N_{c}}[(1-2x)^{2}+1]\frac{1}{\vec{p}_{q\bar{q}}^{\,\,2}}\left\vert
\int\frac{d^{2}p_{\bot}(\vec{p}_{q\bar{q}}\vec{p})}{\vec{p}^{\,\,2}+x\bar
{x}Q^{2}}\mathbf{F}(p_{\bot}+\frac{p_{q\bar{q}\bot}}{2})\right\vert ^{2}.
\label{sigmatt}%
\end{equation}
As a result the convolution of the electron tensor and the photon cross
section reads%
\begin{align}
J^{ab}\left.  \frac{d\sigma_{\gamma^{\ast}p}^{ab}}{dxd\vec{p}_{q}d\vec
{p}_{\bar{q}}}\right\vert _{t=0}  &  =\frac{4Q^{2}}{y^{2}}\left[  \left.
\frac{d\sigma_{0TT}}{dxd\vec{p}_{q}d\vec{p}_{\bar{q}}}\right\vert
_{t=0}\left\{  \frac{1+\bar{y}^{2}}{2}-\frac{x\bar{x}}{1-2x\bar{x}}2\bar
{y}\cos(2\phi)\right\}  \right. \\
&  +\left.  \bar{y}\left.  \frac{d\sigma_{0LL}}{dxd\vec{p}_{q}d\vec{p}%
_{\bar{q}}}\right\vert _{t=0}+\left(  2-y\right)  \sqrt{\bar{y}}\cos
\phi\left.  \frac{d\sigma_{0TL}^{i}e_{i\bot}^{(x)}}{dxd\vec{p}_{q}d\vec
{p}_{\bar{q}}}\right\vert _{t=0}\right]  .
\end{align}
Here $\phi$ is the angle between the quark and the electron's transverse
momenta in our frame. Experimentally $\phi$ is the angle between the jet and
the electron, and the jet may come from the\ antiquark. Then the angle between
the quark and the electron is $\pi-\phi.$ Therefore one measures the sum of
the cross sections with the quark - electron angle equal to $\phi$ and to
$\pi-\phi$. In this sum the interference term $\sigma_{0TL}^{i}$ vanishes,
$\sigma_{0LL}$ and $\sigma_{0TT}$ become twice bigger, and the angle changes
from 0 to $\pi$. Hence starting from here we will omit the $\sigma_{0TL}^{i}$
contribution, understand $\phi$ as the angle between the jet and the electron,
$\phi\in\lbrack0,\pi],$ and double $\sigma_{0LL}$ and $\sigma_{0TT}.$

Next, we have to substitute a model for the hadronic matrix elements
$\mathbf{F}.$ We will use the Golec-Biernat - W\"usthoff (GBW)
\cite{GolecBiernat:1998js}\ parametrization, which was formulated in the
coordinate space. To get the proper normalization we Fourier transform
(\ref{sigmall}) and compare it with Eq.~(4.48) in ref.~\cite{GolecBiernat:2001hab}.
Using%
\begin{equation}
\frac{1}{\vec{l}^{2}+a^{2}}=\int d^{2}r\frac{K_{0}(ar)}{2\pi}e^{-i\vec{l}%
\vec{r}},\quad\mathbf{F}(\vec{k})=\int d\vec{r}e^{-i\vec{k}\vec{r}}F(\vec{r}),
\end{equation}
we have%
\begin{equation}
\left.  \frac{d\sigma_{0LL}}{dxd\vec{p}_{q}d\vec{p}_{\bar{q}}}\right\vert
_{t=0}=\frac{1}{2(2\pi)^{4}}\frac{4\alpha Q_{q}^{2}}{N_{c}}x^{2}\bar{x}%
^{2}Q^{2}\left\vert \int d^{2}r\frac{K_{0}(\sqrt{x\bar{x}}Qr)}{2\pi}%
e^{i\frac{p_{q\bar{q}}{}_{\bot}}{2}\vec{r}}F(\vec{r})\right\vert ^{2}%
\end{equation}
and
\begin{equation}
\left.  \frac{d\sigma_{0LL}}{dt}\right\vert _{t=0}=\frac{1}{2(2\pi)^{4}}%
\frac{4\alpha Q_{q}^{2}}{N_{c}}\pi\int dxQ^{2}x^{2}\bar{x}^{2}\int d^{2}%
rK_{0}(\sqrt{x\bar{x}}Qr)^{2}F(\vec{r})^{2}.
\end{equation}
Comparing it with (4.48) in ref.~\cite{GolecBiernat:2001hab}, the
GBW parametrization of the forward dipole matrix element in our normalization
reads%
\begin{align}
F_{p_{0\bot}p_{0\bot}}(z_{\bot}) &  =\left.  \frac{\langle P^{\prime}%
(p_{0}^{\prime})|T(tr(U_{\frac{z_{\bot}}{2}}U_{-\frac{z_{\bot}}{2}}^{\dag
})-N_{c})|P(p_{0})\rangle}{2\pi\delta(p_{00^{\prime}}^{-})}\right\vert
_{p_{0}\rightarrow p_{0}^{\prime}}\nonumber\\
&  =F(z_{\bot})=N_{c}\sigma_{0}(1-e^{-\frac{z^{2}}{4R_{0}^{2}}}).
\end{align}
Here%
\begin{equation}
\label{R0}
\quad R_{0}=\frac{1}{Q_{0}}\left(  \frac{x_{P}}{a_{0}}\right)  ^{\frac
{\lambda}{2}},
\end{equation}
with
\begin{equation}
\label{def:xP-general}
x_{P}=\frac{Q^{2}+M^2-t}{Q^{2}+W^{2}}.
\end{equation}
which describes the fraction of the incident momentum
lost by the proton or carried by the pomeron. Neglecting the $t-$channel exchanged momentum, we will write
\begin{equation}
\label{def:xP}
x_{P}=\frac{Q^{2}+M^2}{Q^{2}+W^{2}}.
\end{equation}

In the above model,
\begin{equation}
Q_{0}=1~{\rm GeV},\quad\sigma_{0}=23.03\,{\rm mb},\quad\lambda=0.288,\quad a_{0}%
=3.04\ast10^{-4}%
\end{equation}
for 3 active flavours. The nonforward matrix element can be written totally in
the impact parameter space
\begin{equation}
F_{p_{0\bot}p_{0^{\prime}\bot}}(z_{\bot})=\int d\vec{b}e^{-i\vec{b}\vec
{p}_{0^{\prime}0}}F_{\vec{b}}(z_{\bot}).
\end{equation}
Here one can take a simple model \cite{Kowalski:2006hc}\ that the $\vec{b}%
$-dependence factorizes into a Gaussian proton profile%
\begin{equation}
F_{\vec{b}}(z_{\bot})=\frac{1}{2\pi B_{G}}e^{-\frac{b^{2}}{2B_{G}}}%
F_{0}(z_{\bot})=\frac{1}{2\pi B_{G}}e^{-\frac{b^{2}}{2B_{G}}}N_{c}\sigma
_{0}(1-e^{-\frac{z^{2}}{4R_{0}^{2}}})
\end{equation}
with
\begin{equation}
B_{G}=4~{\rm GeV}^{-2}.
\end{equation}
We will need this function in the momentum space%
\[
\mathbf{F}_{p_{0\bot}p_{0^{\prime}\bot}}(p_{\bot})=\int d\vec{z}e^{-i\vec
{z}\vec{p}}\int d\vec{b}e^{-i\vec{b}\vec{p}_{0^{\prime}0}}F_{\vec{b}}(z_{\bot
})
\]%
\begin{equation}
=N_{c}\sigma_{0}\left[  (2\pi)^{2}\delta(\vec{p})-4\pi R_{0}^{2}e^{-R_{0}%
^{2}\vec{p}^{\,\,2}}\right]  e^{-\frac{B_{G}}{2}\vec{p}_{0^{\prime}0}^{\,\,2}%
}=\mathbf{F}(p_{\bot})e^{-\frac{B_{G}}{2}\vec{\tau}^{\,\,2}}%
,\label{F-t-dependence}%
\end{equation}
with
\begin{equation}
\label{def:tau}
\vec{\tau}=\vec{p}_{q}+\vec{p}_{\bar{q}}.
\end{equation}
Therefore in (\ref{sigmall}) and (\ref{sigmatt})
\begin{equation}
\mathbf{F}(p_{\bot})=(2\pi)^{2}N_{c}\sigma_{0}\left[  \delta(\vec{p}%
)-\frac{R_{0}^{2}}{\pi}e^{-R_{0}^{2}\vec{p}^{\,\,2}}\right]  .
\end{equation}
Denoting%
\begin{equation}
\label{def:M}
\vec{M}   =\sqrt{x\bar{x}}\left(\frac{\vec{p}_{q}}{x}-\frac{\vec{p}_{\bar{q}}%
}{\bar{x}}\right),
\end{equation}
we thus have
\begin{equation}
\label{def:jacobian-M-tau}
\frac
{\partial(\vec{M},\vec{\tau})}{\partial(\vec{p}_{q},\vec{p}_{\bar{q}})}%
=\frac{1}{x\bar{x}},
\end{equation}
and
\begin{equation}
\label{def:pq-pqbar-pqqbar}
\vec{p}_{q}   =x\vec{\tau}+\sqrt{x\bar{x}}\,\vec{M},\quad\vec{p}_{\bar{q}}%
=\bar{x}\vec{\tau}-\sqrt{x\bar{x}}\,\vec{M},\quad\vec{p}_{q\bar{q}}=(x-\bar
{x})\vec{\tau}+2\sqrt{x\bar{x}}\,\vec{M}\,.
\end{equation}
One then gets%
\begin{equation}
dxd\vec{p}_{q}d\vec{p}_{\bar{q}}=x\bar{x}dxd\vec{M}d\vec{\tau}=x\bar
{x}dx\,\,\frac{dM^2}{2}d\phi\,\frac{d\tau^{2}}{2}d\phi_{\tau}=\frac{\pi
}{2B_{G}}x\bar{x}dx\,\,dM^2d\phi\,.
\end{equation}
Here $\phi$ is the relative angle between the jet and leptonic planes. 
It is useful to introduce the Bjorken variable $\beta$ normalized to the pomeron momentum, which reads 
\begin{equation}
\label{def:beta-general}
\beta  =\frac{Q^{2}}{Q^{2}+M^2-t}\,.
\end{equation}
Neglecting the $t$-channel exchanged momentum (experimentally, $t$ could not be measured in ZEUS analysis, but was presumably rather small), we will use the simplified expression
\begin{equation}
\label{def:beta}
\beta  =\frac{Q^{2}}{Q^{2}+M^2}\,,
\end{equation}
and thus, denoting $\bar{\beta}=1-\beta,$
\begin{equation}
\label{m2Q2}
 M^2%
=Q^{2}\frac{\bar{\beta}}{\beta}.
\end{equation}
We need the differential 
cross section in $x$, $\beta$ and $\phi$. From
\begin{equation}
\frac{dM^2}{d\beta}=-\frac{Q^{2}}%
{\beta^{2}}, 
\end{equation}
we thus have
\begin{equation}
dxd\vec{p}_{q}d\vec{p}_{\bar{q}}\rightarrow\frac{Q^{2}\pi
}{2\beta^{2}B_{G}}x\bar{x}dx\,\,d\beta d\phi \,.
\end{equation}

\subsection{Experimental cuts}
\label{SubSec:Experimental-cut}

We will now consider the experimental set-up of the ZEUS collaboration. The HERA kinematics is such that $E_{e^-} = 27.5~$GeV and $E_{p} = 920~$GeV, i.e. $\sqrt{s}=318~$GeV.
The phase space covered by the ZEUS collaboration reads \cite{Abramowicz:2015vnu}
\begin{eqnarray}
& W_{\rm min}=90\, {\rm GeV} < W < W_{\rm max}=250 \, {\rm GeV},\quad Q_{\rm min}=5 \, {\rm GeV} <Q,\\
& M_{\rm min}=5 \, {\rm GeV} < M\,, \quad
 x_{P}
<x_{P\max}=0.01.
\end{eqnarray}
Hence, using eq.~(\ref{def:xP}) one has
\begin{equation}
M^2<x_{P\max}W^{2}-Q^{2}(1-x_{P\max}).
\end{equation}
For fixed $\beta$ we have, using eq.~(\ref{m2Q2}),
\begin{equation}
\label{inequalityQ2}
\max\left(Q_{\rm min}^{2},\frac{\beta}{\bar{\beta}}M_{\rm min}^{2}\right)<Q^{2}<\frac{x_{P {\rm max}}\beta}{1-x_{P {\rm max}}\beta
}W^{2}.
\end{equation}
A careful study shows that
\begin{equation}
\label{beta_min-beta_max}
\beta_{\min}=\frac{Q_{\rm min}^2}{x_{P {\rm max}}(W_{\rm max}^{2}+Q_{\rm min}^2)} \,, \quad
 \beta_{\max}=\frac{W_{\rm max}^{2}\, x_{P {\rm max}}-M_{\rm min}^2}{x_{P {\rm max}}(W_{\rm max}^{2}-M_{\rm min}^2)}\,.
\end{equation}
On the other hand, eq.~(\ref{inequalityQ2}) leads to
\begin{equation}
\label{inequalityW2}
\max\left[W_{\rm min}^{2},\max\left(Q_{\rm min}^2,M_{\rm min}^2\frac{\beta}{\bar{\beta}}\right)\frac{1-x_{P {\rm max}}\beta
}{x_{P {\rm max}}\beta}\right]<W^{2}<W_{\rm max}^{2}\,.
\end{equation}
The inelasticity restriction reads%
\begin{equation}
\label{yconstraint}
y_{\rm min}=0.1<y< y_{\rm max}=0.65, \quad {\rm i.e.} \quad y_{\rm min} s<Q^{2}+W^{2}< y_{\rm max} s\,.
\end{equation}
Eqs.~(\ref{inequalityQ2}) and (\ref{yconstraint}) thus result in the following constraints for $Q^2$:
\begin{equation}
\label{Q2constraint-general}
\max\left(Q_{\rm min}^2,M_{\rm min}^2\frac{\beta}{\bar{\beta}}, y_{\rm min} s-W^{2}\right)<Q^{2}< \min\left(\frac
{x_{P {\rm max}}\beta}{1-x_{P {\rm max}}\beta}W^{2},y_{\rm max} s -W^2\right)
\end{equation}
One should note that in eq.~(\ref{Q2constraint-general}), 
\begin{equation}
\min\left(\frac
{x_{P {\rm max}}\beta}{1-x_{P {\rm max}}\beta}W^{2},y_{\rm max} s -W^2\right)=y_{\rm max} s -W^2
\end{equation}
would mean that
\begin{equation}
\frac{1}{x_{P {\rm max}}}\left(1 - \frac{W^2}{y_{\rm max} s} \right) < \beta <\beta_{\rm max}
\end{equation}
and thus, using the expression of $\beta_{\rm max}$, see eq.~(\ref{beta_min-beta_max}), that 
\begin{equation}
 y_{\rm max} s < W_{\rm max}^{2}+ Q_{\rm min}^2\,.
\end{equation}
For the experimental values of ZEUS, this is not satisfied, and one can thus simplify the constraints (\ref{Q2constraint-general}) on $Q^2$ as
\begin{equation}
\label{Q2constraint}
\max\left(Q_{\rm min}^2,M_{\rm min}^2\frac{\beta}{\bar{\beta}}, y_{\rm min} s-W^{2}\right)<Q^{2}< \frac
{x_{P {\rm max}}\beta}{1-x_{P {\rm max}}\beta}W^{2}\,.
\end{equation}
Similarly, Eqs.~(\ref{inequalityQ2}) and (\ref{yconstraint})  result in the following constraints for $W^2$:
\begin{equation}
\label{W2constraint}
(1-x_{P {\rm max}}\beta)\max\left(y_{\rm min} s,\frac{Q_{\rm min}^2}{x_{P {\rm max}}\beta},\frac{M_{\rm min}^2}{x_{P {\rm max}}\bar{\beta
}}\right)<W^{2}<W_{\rm max}^{2}\,.
\end{equation}
Additionally, there is a restriction on the transverse momentum of the jet
\begin{equation}
\label{p-constraint}
p>p_{\min}=2\, {\rm GeV}.
\end{equation}
In the $t=0$ limit, i.e. $\vec{\tau}=\vec{0}$, we have from eq.~(\ref{def:pq-pqbar-pqqbar}) $p = |\vec{p}_q| = |\vec{p}_{\bar{q}}|= \sqrt{x \bar{x}} M$. 
Thus, the contraint (\ref{p-constraint}) reads
\begin{equation}
p^2 = x \bar{x} \, M^2 = x \bar{x} \frac{\bar{\beta}}{\beta} Q^2 > p_{\rm min}^2
\end{equation}
and leads to the following
restrictions on $x$:
\begin{equation}
x\in\lbrack x_{\min},\bar{x}_{\min}]\equiv\lbrack\frac{1}{2}-\sqrt{\frac{1}%
{4}-\frac{p_{\min}^{2}\beta}{Q^{2}\bar{\beta}}},\frac{1}{2}+\sqrt{\frac{1}%
{4}-\frac{p_{\min}^{2}\beta}{Q^{2}\bar{\beta}}}].\label{xmin-xmax}%
\end{equation}

There is one more experimental cut imposed in ref.~\cite{Abramowicz:2015vnu}. It is
the restriction on the jet rapidity $\eta_{\max}=2$, where the rapidity is
defined in the detector frame with the $z$ axis along the proton and electron
velocities in the proton beam direction. One can rewrite this cut as cut on
$x_{min}$ as well. Indeed, one can transform momenta from the proton-photon frame
(\ref{proton-photon-frame1}--\ref{proton-photon-frame-2}) to the detector
frame. For any vector $l$%
\begin{equation}
l_{Det}=R_{xz}\Lambda_{x}R_{xy}\Lambda_{z}l,
\end{equation}%
\begin{align}
\Lambda &  :l^{+}\rightarrow\lambda l^{+},l^{-}\rightarrow\frac{1}{\lambda
}l^{-},\quad R_{xy}=%
\begin{pmatrix}
1 &  &  & \\
& \cos\phi & \sin\phi & \\
& -\sin\phi & \cos\phi & \\
&  &  & 1
\end{pmatrix}
,\\
\Lambda_{x} &  =%
\begin{pmatrix}
\gamma & \beta\gamma &  & \\
\beta\gamma & \gamma &  & \\
&  & 1 & \\
&  &  & 1
\end{pmatrix}
,\quad R_{xz}=%
\begin{pmatrix}
1 &  &  & \\
& \cos\alpha &  & -\sin\alpha\\
&  & 1 & \\
& \sin\alpha &  & \cos\alpha
\end{pmatrix}
,
\end{align}
where%
\begin{equation}
\beta=-\sin\alpha,\quad\gamma=\frac{1}{\cos\alpha},\quad\sin\alpha=\frac
{k}{2\lambda p_{e}^{+}}.
\end{equation}
After this transformation one gets%
\begin{align}
p_{0Det} &  =\frac{sy}{\sqrt{(2\lambda yp_{e}^{+})^{2}-Q^{2}\bar{y}}}%
n_{2},\quad k_{Det}=\frac{1}{2y}\sqrt{(2\lambda yp_{e}^{+})^{2}-Q^{2}\bar{y}%
}n_{1},\\
q_{Det} &  =\frac{1}{2}\sqrt{(2\lambda yp_{e}^{+})^{2}-Q^{2}\bar{y}}%
n_{1}-\frac{Q^{2}y}{\sqrt{(2\lambda yp_{e}^{+})^{2}-Q^{2}\bar{y}}}n_{2}%
-Q\sqrt{\overline{y}}e_{x\bot},\\
p_{qDet} &  =\frac{x}{2}\sqrt{(2\lambda yp_{e}^{+})^{2}-Q^{2}\bar{y}}%
n_{1}+\frac{\vec{p}_{q}^{\,\,2}-2xp_{q}Q\sqrt{\overline{y}}\cos\phi+Q^{2}%
x^{2}\overline{y}}{x\sqrt{(2\lambda yp_{e}^{+})^{2}-Q^{2}\bar{y}}}\nonumber\\
&  +(p_{q}\cos\phi-Qx\sqrt{\overline{y}})e_{x\bot}-p_{q}\sin\phi e_{y\bot}.
\end{align}
In the detector frame
\begin{equation}
k_{Det}^{+}=\frac{1}{2y}\sqrt{(2\lambda yp_{e}^{+})^{2}-Q^{2}\bar{y}}%
=E_{e}=27.5\,GeV.
\end{equation}
This condition fixes $p_{e}^{+}$ or $\lambda$, the remaining parameter
representing freedom in $z$-boosts in the $\gamma$-proton frame. Then
$p_{qDet}$'s rapidity reads%
\begin{equation}
\eta_{qDet}=\frac{1}{2}\ln\frac{4E_{e}^{2}x^{2}y^{2}}{\vec{p}_{q}%
^{\,\,2}-2xp_{q}Q\sqrt{\overline{y}}\cos\phi+Q^{2}x^{2}\overline{y}}%
>-\eta_{\max},\label{rapiditycut1}%
\end{equation}
where we changed the sign to 
 take into account the propagation along the negative $z$ direction
 (the $z$ axis in the ZEUS frame and in our frame are opposite). Obviously, this constraint should be fulfilled for both quark and the antiquark jets, i.e. eq.~(\ref{rapiditycut1}) with $x \to \bar{x}.$ A careful inspection then shows that these two constraints turn into  
\begin{align}
&  x,\bar{x}>x_{0}=\bar{\beta}\label{rapiditycut2}\\
&  \!\!\times \! \frac{\beta\left( \! \frac{2E_{e}y}{Q}\!\right)^{2}\!\! -2\beta
e^{-\eta_{\max}}\cos\phi\sqrt{\bar{y}\!\left( \!\!\left(  \frac{2E_{e}y}{Q}\right)
^{2}\!-e^{-2\eta_{\max}}\bar{y}\sin^{2}\phi\!\right)  }\!+e^{-2\eta_{\max}}\!\left(
\bar{\beta}+\beta\bar{y}\cos(2\phi)\right)  }{2\beta\bar{y}\!\left(
\!e^{-2\eta_{\max}}\bar{\beta}\cos(2\phi)\!-\!\beta\left(  \frac{2E_{e}y}{Q}\right)
^{2}\right) \!\! +e^{2\eta_{\max}}\!\left(  e^{-2\eta_{\max}}\bar{\beta}%
+\beta\left(  \frac{2E_{e}y}{Q}\right)^{2}\right)^2\!\!+\beta^{2}%
e^{-2\eta_{\max}}\bar{y}^{2}}.\nonumber
\end{align}
The minimal value for $x$ is thus
\begin{equation}
\tilde{x}_{\min}=\max(x_{\min},x_{0}),
\end{equation}
with the additional constraint that $\tilde{x}_{\min}<\frac{1}{2}.$ However as we will show later, numerically
this rapidity restriction is negligible. Therefore we will include it only
in the discussion of the final result. 

Finally, one has to calculate
\begin{align}
\frac{d\sigma_{ep}}{d\beta d\phi} &  =2\frac{\alpha}{\beta^{2}B_{G}}
\int_{(1-x_{P {\rm max}}\beta)\max\left(y_{\rm min}s,\frac{Q_{\rm min}^2}{x_{P {\rm max}}\beta},\frac{M_{\rm min}^2}%
{x_{P {\rm max}}\bar{\beta}}\right)}^{W_{\rm max}^{2}}\!\frac{dW^{2}}{2s}\int_{\max\left(Q_{\rm min}^2,M_{\rm min}^2%
\frac{\beta}{\bar{\beta}},y_{\rm min}s-W^{2}\right)}^{\frac{x_{P {\rm max}}\beta}{1-x_{P {\rm max}}\beta}W^{2}%
}\!\frac{dQ^{2}}{y}\nonumber\\
\times &  \int_{x_{\min}}^{\bar{x}_{\min}}x\bar{x}dx\left[  \bar{y}\left.
\frac{d\sigma_{0LL}}{dxd\vec{p}_{q}d\vec{p}_{\bar{q}}}\right\vert
_{t=0}\right.  \left.  +\left.  \frac{d\sigma_{0TT}}{dxd\vec{p}_{q}d\vec
{p}_{\bar{q}}}\right\vert _{t=0}\left\{  \frac{1+\bar{y}^{2}}{2}-\frac
{2\bar{y}x\bar{x}}{1-2x\bar{x}}\cos(2\phi)\right\}  \right]  ,\label{dsigmaLO}%
\end{align}
where $\phi\in\lbrack0,\pi]$ and
\begin{align}
\hspace{-.1cm}\left.  \frac{d\sigma_{0LL}}{dxd\vec{p}_{q}d\vec{p}_{\bar{q}}}\right\vert
_{t=0}\!\!\!\!\! &  =\frac{1}{(\hbar c)^{2}}\frac{1}{2(2\pi)^{4}}\frac{4\alpha Q_{q}%
^{2}}{(2\pi)^{4}N_{c}}x^{2}\bar{x}^{2}Q^{2}\left\vert \int\!\!\frac{d^{2}p_{\bot}%
}{\vec{p}^{\,\,2}+x\bar{x}Q^{2}}\mathbf{F}(p_{\bot}+\frac{p_{q\bar{q}}{}%
_{\bot}}{2})\right\vert ^{2},\label{sigmaloL}\\
\hspace{-.1cm}\left.  \frac{d\sigma_{0TT}}{dxd\vec{p}_{q}d\vec{p}_{\bar{q}}}\right\vert
_{t=0}\!\!\!\!\! &  =\frac{1}{(\hbar c)^{2}}\frac{1}{2(2\pi)^{4}}\frac{1}{2}\frac{\alpha
Q_{q}^{2}}{(2\pi)^{4}N_{c}}[(1-2x)^{2}+1]\frac{1}{\vec{p}_{q\bar{q}}^{\,\,2}%
}\left\vert \int\!\!\frac{d^{2}p_{\bot}(\vec{p}_{q\bar{q}}\vec{p})}{\vec
{p}^{\,\,2}+x\bar{x}Q^{2}}\mathbf{F}(p_{\bot}+\frac{p_{q\bar{q}\bot}}%
{2})\right\vert ^{2},\label{sigmaloT}%
\end{align}
with%
\begin{align}
\vec{p}_{q\bar{q}} &  =2\sqrt{x\bar{x}}m\,\vec{e}^{(x)}=2\sqrt{x\bar{x}%
\frac{\bar{\beta}}{\beta}}Q\,\vec{e}^{(x)},\\
\mathbf{F}(p_{\bot}) &  =(2\pi)^{2}N_{c}\sigma_{0}\left[  \delta(\vec
{p})-\frac{R_{0}^{2}}{\pi}e^{-R_{0}^{2}\vec{p}^{\,\,2}}\right]  =-(2\pi
)^{2}\frac{N_{c}\sigma_{0}}{\pi}e^{-R_{0}^{2}\vec{p}^{\,\,2}}\frac{\partial
}{\partial p^{2}}.
\end{align}
The $t$-channel integrals can be simplified%
\begin{align}
&  \int\frac{d^{2}p_{\bot}}{\vec{p}^{\,\,2}+x\bar{x}Q^{2}}\mathbf{F}(p_{\bot
}+\frac{p_{q\bar{q}}{}_{\bot}}{2})=\int_{0}^{+\infty}\frac{\pi\mathbf{F}%
(p)dp^{2}}{\sqrt{(x\bar{x}Q^{2}+p^{2}+(\frac{\vec{p}_{q\bar{q}}{}}{2}%
)^{2})^{2}-4p^{2}(\frac{\vec{p}_{q\bar{q}}{}}{2})^{2}}}\\
&  =-N_{c}\sigma_{0}\int_{0}^{+\infty}dp^{2}e^{-R_{0}^{2}\vec{p}^{\,\,2}}%
\frac{\partial}{\partial p^{2}}\frac{(2\pi)^{2}}{\sqrt{(x\bar{x}Q^{2}%
+p^{2}+(\frac{\vec{p}_{q\bar{q}}{}}{2})^{2})^{2}-4p^{2}(\frac{\vec{p}%
_{q\bar{q}}{}}{2})^{2}}},\label{loLifExactGBW}%
\end{align}%
\begin{align}
&  \frac{1}{\left\vert \vec{p}_{q\bar{q}}\right\vert }\int\frac{d^{2}p_{\bot
}(\vec{p}_{q\bar{q}}\vec{p})}{\vec{p}^{\,\,2}+x\bar{x}Q^{2}}\mathbf{F}%
(p_{\bot}+\frac{p_{q\bar{q}\bot}}{2})=\int_{0}^{+\infty}\frac{dp^{2}}%
{2}\mathbf{F}(p)\int d\phi\frac{(\vec{e}^{(x)},\vec{p}-\frac{\vec{p}_{q\bar
{q}}}{2})}{(\vec{p}-\frac{\vec{p}_{q\bar{q}}}{2})^{\,\,2}+x\bar{x}Q^{2}}\\
&  =\pi\int_{0}^{+\infty}dp^{2}\frac{\mathbf{F}(p)}{2\frac{\left\vert \vec
{p}_{q\bar{q}}\right\vert }{2}}\left(  \frac{x\bar{x}Q^{2}+p^{2}-(\frac
{\vec{p}_{q\bar{q}}{}}{2})^{2}}{\sqrt{(x\bar{x}Q^{2}+p^{2}+(\frac{\vec
{p}_{q\bar{q}}{}}{2})^{2})^{2}-4p^{2}(\frac{\vec{p}_{q\bar{q}}{}}{2})^{2}}%
}-1\right)  \\
&  =-(2\pi)^{2}\frac{N_{c}\sigma_{0}}{2\frac{\left\vert \vec{p}_{q\bar{q}%
}\right\vert }{2}}\int_{0}^{+\infty}dp^{2}e^{-R_{0}^{2}\vec{p}^{\,\,2}}%
\frac{\partial}{\partial p^{2}}\frac{x\bar{x}Q^{2}+p^{2}-(\frac{\vec{p}%
_{q\bar{q}}{}}{2})^{2}}{\sqrt{(x\bar{x}Q^{2}+p^{2}+(\frac{\vec{p}_{q\bar{q}}%
{}}{2})^{2})^{2}-4p^{2}(\frac{\vec{p}_{q\bar{q}}{}}{2})^{2}}}%
.\label{loTifExactGBW}%
\end{align}
These integrals will be calculated numerically.

\subsection{BFKL-like approximation}
\label{SubSec:BFKL-Approximation}

\begin{figure}[ptb]%
\centering
\raisebox{7cm}{\includegraphics[angle=-90,
width = 140 mm,
]%
{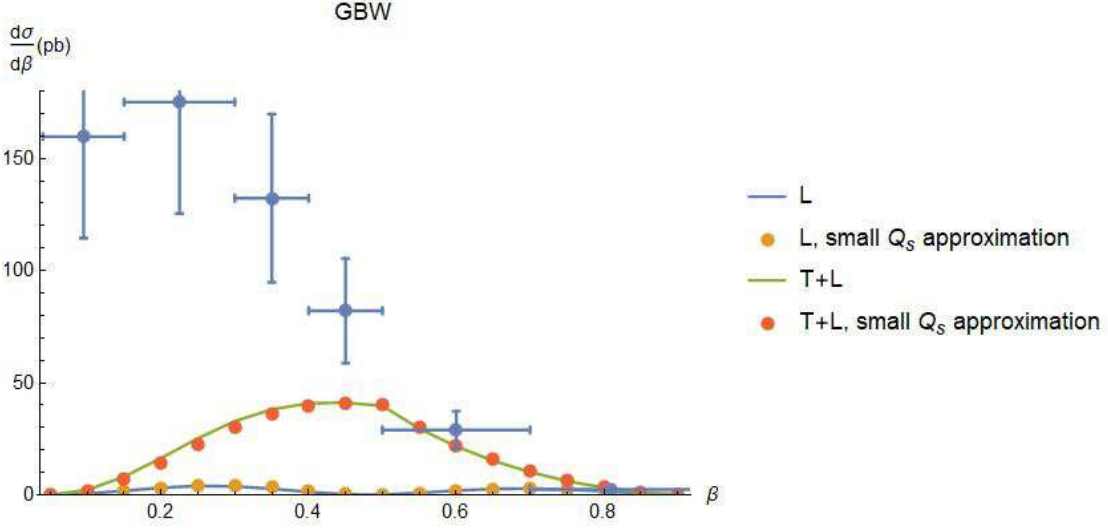}}%
\caption{Longitudinal (L) and both transverse and longitudinal photon
contributions to the dijet cross section calculated exactly from LO formulas
(\ref{loLifExactGBW}-\ref{loTifExactGBW}) and in the small $Q_{s}$ approximation
(\ref{loLifSmallQsGBW}-\ref{loTifSmallQsGBW}).}%
\label{loExactSmallQs}%
\end{figure}

In our kinematics the saturation scale is much lower than all other
scales. Indeed, we have%
\begin{equation}
\frac{Q_{\rm min}^2+M_{\rm min}^2}{Q_{\rm min}^2+W_{\rm max}^{2}}\simeq0.0008<x_{P}<0.01, \label{smallQs1}%
\end{equation}%
\begin{equation}
Q_{s}^{2}=\frac{1}{R_{0}^{2}}<0.8\,{\rm GeV}^{2}<p_{\min}^{2}=2^{2}{\rm GeV}^{2}\ll
Q^{2},M^{2}\in\lbrack5^{2},25^{2}]\,GeV^{2}. \label{smallQs2}%
\end{equation}
It means that neglecting $p^{2}$ in the denominator in (\ref{loTifExactGBW})
gives the error
\begin{equation}
\sim\left\vert \frac{2p^{2}(x\bar{x}Q^{2}-(\frac{\vec{p}_{q\bar{q}}{}}{2}%
)^{2})}{(x\bar{x}Q^{2}+(\frac{\vec{p}_{q\bar{q}}{}}{2})^{2})^{2}}\right\vert
\leq\frac{2p^{2}}{x\bar{x}Q^{2}+p_{\min}^{2}}.
\end{equation}
Therefore at least with $O(\frac{Q_{s}^{2}}{p_{\min}^{2}})$ precision one can
neglect the $t$-channel momentum in the integrals and calculate them analytically
to get%
\begin{eqnarray}
&&\int\frac{d^{2}p_{\bot}}{\vec{p}^{\,\,2}+x\bar{x}Q^{2}}\mathbf{F}(p_{\bot
}+\frac{p_{q\bar{q}}{}_{\bot}}{2})\simeq-(2\pi)^{2}\frac{N_{c}\sigma_{0}%
}{R_{0}^{2}}\frac{(\frac{\vec{p}_{q\bar{q}}{}}{2})^{2}-x\bar{x}Q^{2}}%
{((\frac{\vec{p}_{q\bar{q}}{}}{2})^{2}+x\bar{x}Q^{2})^{3}} \nonumber \\
&&
=-\frac{(2\pi)^{2}N_{c}\sigma_{0}}{R_{0}^{2}(x\bar{x})^{2}}\frac{M^2-Q^{2}%
}{(M^2+Q^{2})^{3}}=-(2\pi)^{2}\frac{N_{c}\sigma_{0}(\bar{\beta}-\beta
)\beta^{2}}{Q^{4}R_{0}^{2}(x\bar{x})^{2}}, \label{loLifSmallQsGBW}%
\end{eqnarray}%
and
\begin{eqnarray}
&&\int\frac{d^{2}p_{\bot}(\vec{p}_{q\bar{q}}\vec{p})}{\vec{p}^{\,\,2}+x\bar
{x}Q^{2}}\frac{\mathbf{F}(p_{\bot}+\frac{p_{q\bar{q}\bot}}{2})}{\left\vert
\vec{p}_{q\bar{q}}\right\vert }\simeq-\frac{N_{c}\sigma_{0}}{R_{0}^{2}}%
\frac{(2\pi)^{2}2\frac{\left\vert \vec{p}_{q\bar{q}}\right\vert }{2}x\bar
{x}Q^{2}}{((\frac{\vec{p}_{q\bar{q}}{}}{2})^{2}+x\bar{x}Q^{2})^{3}}%
\nonumber \\
&&
=-\frac{(2\pi)^{2}N_{c}\sigma_{0}}{R_{0}^{2}(x\bar{x})^{\frac{3}{2}}}%
\frac{2mQ^{2}}{(M^2+Q^{2})^{3}}=-(2\pi)^{2}\frac{N_{c}\sigma_{0}2\sqrt
{\bar{\beta}}\beta^{\frac{5}{2}}}{Q^{3}R_{0}^{2}(x\bar{x})^{\frac{3}{2}}}.
\label{loTifSmallQsGBW}%
\end{eqnarray}
In this approximation the $ep$ cross section (\ref{dsigmaLO}) reads%
\begin{eqnarray}
\label{dsigmaBorn}
&&\left.\frac{d\sigma_{ep}}{d\beta d\phi}\right|_{\rm Born}   =\frac{2\alpha^{2}Q_{q}^{2}%
N_{c}\sigma_{0}^{2}}{(2\pi)^{4}B_{G}(\hbar c)^{2}}\int_{(1-x_{P {\rm max}}\beta
){\rm max}\left(y_{\rm min}s,\frac{Q_{\rm min}^2}{x_{P {\rm max}}\beta},\frac{M_{\rm min}^2}{x_{P {\rm max}}\bar{\beta}}\right)}^{W_{\rm max}^{2}%
}\frac{dW^{2}}{s} \\
&&\times \int_{\max\left(Q_{\rm min}^2,M_{\rm min}^2\frac{\beta}{\bar{\beta}},y_{\rm min}s-W^{2}%
\right)}^{\frac{x_{P {\rm max}}\beta}{1-x_{P {\rm max}}\beta}W^{2}}\frac{dQ^{2}}{yR_{0}^{4}Q^{6}%
}\nonumber\\
&&  \times\int_{x_{\min}}^{\bar{x}_{\min}}dx\left[  \bar{y}\frac{(\bar{\beta
}-\beta)^{2}\beta^{2}}{x\bar{x}}+\frac{[(1-2x)^{2}+1]}{2}\frac{\bar{\beta
}\beta^{3}}{(x\bar{x})^{2}}\left\{  \frac{1+\bar{y}^{2}}{2}-\frac{2\bar
{y}x\bar{x}}{1-2x\bar{x}}\cos(2\phi)\right\}  \right]  . \nonumber
\end{eqnarray}
Then  the integral w.r.t. $x$ can be performed analytically
\begin{eqnarray}
\label{dsigma/dbetaddfiAnalytic}
&&\frac{d\sigma_{ep}}{d\beta d\phi}|_{Born}    =\frac{4\alpha^{2}Q_{q}^{2}%
N_{c}\sigma_{0}^{2}}{(2\pi)^{4}B_{G}(\hbar c)^{2}}\int_{(1-x_{P {\rm max}}\beta
)\max\left(y_{\rm min}s,\frac{Q_{\rm min}^2}{x_{P {\rm max}}\beta},\frac{M_{\rm min}^2}{x_{P {\rm max}}\bar{\beta}}\right)}^{W_{\rm max}^{2}%
}\frac{dW^{2}}{s} \\
&&\times \int_{\max\left(Q_{\rm min}^2,M_{\rm min}^2\frac{\beta}{\bar{\beta}},y_{\rm min}s-W^{2}%
\right)}^{\frac{x_{P {\rm max}}\beta}{1-x_{P {\rm max}}\beta}W^{2}}\frac{dQ^{2}}{yR_{0}^{4}Q^{6}%
}\nonumber\\
&&  \times\left[  \bar{y}(\bar{\beta}-\beta)^{2}\beta^{2}\ln\left(  \frac
{\bar{x}_{\min}}{x_{\min}}\right)  +\bar{\beta}\beta^{3}\left\{  \frac
{1+\bar{y}^{2}}{2}\frac{1-2x_{\min}}{x_{\min}\bar{x}_{\min}}-2\bar{y}%
\ln\left(  \frac{\bar{x}_{\min}}{x_{\min}}\right)  \cos(2\phi)\right\}
\right]  . \nonumber%
\end{eqnarray}
The results integrated w.r.t. $\phi\in\lbrack0,\pi]$ are in figure
\ref{loExactSmallQs}. As one can see the approximation errors are smaller than
the experimental ones.

\subsection{Analysis of the LO result}
\label{SubSec:LO-analysis}

Following ref.~\cite{GolecBiernat:2001hab},\ we rewrite (\ref{dsigma/dbetaddfiAnalytic})
in terms of diffractive structure functions $F^{D}$. These functions are
defined through%
\begin{equation}
\frac{d\sigma_{ep}}{dx_{B}dQ^{2}dx_{P}dt}=\frac{4\pi\alpha^{2}}{x_{B}Q^{4}%
}\left\{  \frac{1+\bar{y}^{2}}{2}F_{T}^{D(4)}+\bar{y}F_{L}^{D(4)}\right\}  ,
\end{equation}%
\begin{equation}
F_{T,L}^{D(4)}(x_{B},Q^{2},x_{P},t)=\frac{dF_{T,L}^{D}}{dx_{P}dt}=\frac{Q^{2}%
}{4\pi^{2}\alpha}\frac{d\sigma_{T,L}^{\gamma p}}{dx_{P}dt},\quad
F_{T,L}^{D(3)}=\int_{-\infty}^{0}dtF_{T,L}^{D(4)},
\end{equation}
where
\begin{equation}
x_{B}=\frac{Q^{2}}{2p_{0}q}=\frac{Q^{2}}{W^{2}+Q^{2}}=\beta x_{P}%
\end{equation}
is the Bjorken variable. Since%
\begin{equation}
dx_{B}=-\frac{x_{B}^{2}}{Q^{2}}dW^{2},\quad dx_{P}=-x_{P}\frac{d\beta}{\beta},
\end{equation}
one gets%
\begin{equation}
x_{P}F_{T}^{D(3)}=\frac{Q_{q}^{2}N_{c}\sigma_{0}^{2}}{(2\pi)^{4}B_{G}(\hbar
c)^{2}}\frac{\bar{\beta}\beta^{4}}{Q^{2}R_{0}^{4}}\frac{1-2x_{\min}}{x_{\min
}\bar{x}_{\min}},
\end{equation}%
\begin{equation}
x_{P}F_{L}^{D(3)}=\frac{Q_{q}^{2}N_{c}\sigma_{0}^{2}}{(2\pi)^{4}B_{G}(\hbar
c)^{2}}\frac{(\bar{\beta}-\beta)^{2}\beta^{3}}{Q^{2}R_{0}^{4}}\ln\left(
\frac{\bar{x}_{\min}}{x_{\min}}\right)  ,
\end{equation}
which gives in the small $\beta$ ($M^2\gg Q^{2}$) region%
\begin{equation}
x_{P}F_{T}^{D(3)}\simeq\frac{\sigma_{0}^{2}}{B_{G}}\frac{\beta^{4}}{Q^{2}%
R_{0}^{4}}\frac{Q^{2}}{p_{\min}^{2}\beta},\quad x_{P}F_{L}^{D(3)}\simeq
\frac{\sigma_{0}^{2}}{B_{G}}\frac{\beta^{3}}{Q^{2}R_{0}^{4}}\ln\left(
\frac{Q^{2}}{p_{\min}^{2}\beta}\right)  .
\end{equation}
This behavior contradicts the known one \cite{GolecBiernat:2001hab}%
\begin{equation}
x_{P}\tilde{F}_{T}^{D(3)}\sim\frac{\sigma_{0}^{2}\beta\bar{\beta}}{B_{G}%
R_{0}^{2}},\quad x_{P}\tilde{F}_{L}^{D(3)}\sim\frac{\sigma_{0}^{2}}{B_{G}%
}\frac{\beta^{3}}{Q^{2}R_{0}^{4}},\label{GBstructureFunctions}%
\end{equation}
where we introduced $\tilde{F}$ to distinguish them from our result.

First, let us emphasize that our transverse structure function $F_{T}^{D(3)}$ is correctly
proportional to $\bar{\beta}.$ Indeed, since the final $q\bar{q}$ pair has
opposite helicities, it carries angular momentum as orbital momentum and its
wave function scales like $p_{\bot}\sim M.$ Therefore it should vanish at $M=0$, i.e.
$\beta=1.$

Next, $F_{T}^{D(3)}$ is a higher twist correction compared to
(\ref{GBstructureFunctions}) as it has an extra power of $Q^{2}R_{0}^{2}\gg1$
in its denominator. The origin of this suppression lies in the fact that the
dominant contribution to the transverse cross section comes from the aligned
jet configuration, i.e. the region of $x\lesssim\frac{1}{\max(Q^{2}%
,M^2)R_{0}^{2}}\ll1$. We discuss it in Appendix~\ref{append:Scaling}. However in our
kinematics\ (\ref{xmin-xmax}), (\ref{smallQs1}--\ref{smallQs2})
\begin{equation}
\frac{1}{M^2R_{0}^{2}}<\frac{0.8\,GeV^{2}}{M^2}\ll x_{\min}=\frac{1}%
{2}-\sqrt{\frac{1}{4}-\frac{4\,GeV^{2}}{M^2}},
\end{equation}
which for the largest $M^2\simeq25^{2}$ GeV$^{2}$ gives%
\begin{equation}
\frac{1}{M^2R_{0}^{2}}\simeq0.001\ll x_{\min}\simeq0.006.
\end{equation}
Therefore the current experimental setup does not let us probe the leading
twist contribution to the transverse cross section. In other words the
experimental cuts kill the leading twist aligned jets which come from the
saturation region. As a result we are left with the subleading twist
perturbative BFKL-like ($\sigma\sim s^{2\lambda}$) behavior
(\ref{dsigma/dbetaddfiAnalytic}). One can also feel that the experiment sees
only the subleading twist contribution from fig. 6d in
ref.~\cite{Abramowicz:2015vnu} where they cut off the $p_{\bot}$ distribution peak.

The longitudinal structure function is subleading to the transverse one in
twist (\ref{GBstructureFunctions}). The whole $0<x<1$ range contributes to it. Therefore the experimental cuts only change the $\beta$-dependence
of the result.

\section{$k_{t}$ jet algorithm}
\label{Sec:jet-algorithm}

\subsection{Exclusive $k_{t}$ jet algorithm for three partons}
\label{SubSec:jet-algorithm-3p}

Let us recall the parametrization of the momenta of the 3 outgoing partons.
For the 3 particles with the momenta
\begin{align}
p_{q}  &  =x_{q}p_{\gamma}^{+}n_{1}+\frac{\vec{p}_{q}^{\,\,2}}{2x_{q}%
p_{\gamma}^{+}}n_{2}+p_{q\bot},\quad p_{\bar{q}}=x_{\bar{q}}p_{\gamma}%
^{+}n_{1}+\frac{\vec{p}_{\bar{q}}^{\,\,2}}{2x_{\bar{q}}p_{\gamma}^{+}}%
n_{2}+p_{\bar{q}\bot},\\
p_{g}  &  =zp_{\gamma}^{+}n_{1}+\frac{\vec{p}_{g}^{\,\,2}}{2zp_{\gamma}^{+}%
}n_{2}+p_{g\bot},\quad\quad p=p_{q}+p_{\bar{q}}+p_{g},\quad M^2=p^{2},
\end{align}
in the c.m.f.%
\begin{equation}
p=p_{\gamma}^{+}n_{1}+\frac{M^2}{2p_{\gamma}^{+}}n_{2},\quad p_{\gamma}%
^{+}=\frac{M}{2}.
\end{equation}
The distance between two particles according to the $k_{t}$ algorithm
\cite{Brown:1991hx} reads
\begin{equation}
d_{ij}=2\min(E_{i}^{2},E_{j}^{2})\frac{1-\cos\theta_{ij}}{M^2}=\min
\left(\frac{E_{i}}{E_{j}},\frac{E_{j}}{E_{i}}\right)\frac{2p_{i}p_{j}}{M^2}=\min
\left(\frac{p_{i}p}{p_{j}p},\frac{p_{j}p}{p_{i}p}\right)\frac{2p_{i}p_{j}}{M^2}.
\end{equation}
Here $E_{i,j},$ $\theta_{ij}$ are the particle's energies and the relative
angle between them in c.m.f. Two particles belong to one jet if $d_{ij}%
<y_{cut}.$ In our case $y_{cut}=0.15$ \cite{Abramowicz:2015vnu}.

One introduces the variables
\begin{equation}
\mathbf{x}_{i}=2\frac{E_{i}}{M}=\frac{2pp_{i}}{M^2}\leq1,
\end{equation} 
which satisfy
\begin{equation}
\sum
_{i=q,\bar{q},g}\mathbf{x}_{i}=2,\ \frac{2p_{i}p_{j}}{M^2}=1-\mathbf{x}%
_{k}=\mathbf{\bar{x}}_{k},\ d_{ij}=\min\left(\frac{\mathbf{x}_{i}}%
{\mathbf{x}_{j}},\frac{\mathbf{x}_{j}}{\mathbf{x}_{i}}\right)(1-\mathbf{x}_{k}).
\end{equation}
In our variables%
\begin{align}
\frac{(x_{\bar{q}}\vec{p}_{q}-x_{q}\vec{p}_{\bar{q}})^{2}}{x_{q}x_{\bar{q}%
}M^2}  &  =1-\mathbf{x}_{g}=\frac{(p-p_{g})^{2}}{M^2}=\frac{(p_{q}%
+p_{\bar{q}})^{2}}{M^2}=\frac{2p_{q}p_{\bar{q}}}{M^2},\label{mass1}\\
\frac{(x_{\bar{q}}\vec{p}_{g}-z\vec{p}_{\bar{q}})^{2}}{zx_{\bar{q}}M^2}  &
=1-\mathbf{x}_{q},\quad\frac{(z\vec{p}_{q}-x_{q}\vec{p}_{g})^{2}}{x_{q}zM^2%
}=1-\mathbf{x}_{\bar{q}},
\end{align}
and using
\begin{equation}
\bar{\mathbf{x}}_{q}+\bar{\mathbf{x}}_{\bar{q}}+\bar{\mathbf{x}}_{g}=1
\end{equation}
we have
\begin{align}
M^2  &  =\frac{(x_{\bar{q}}\vec{p}_{g}-z\vec{p}_{\bar{q}})^{2}}{zx_{\bar{q}%
}}+\frac{(z\vec{p}_{q}-x_{q}\vec{p}_{g})^{2}}{x_{q}z}+\frac{(x_{\bar{q}}%
\vec{p}_{q}-x_{q}\vec{p}_{\bar{q}})^{2}}{x_{q}x_{\bar{q}}}. \label{mass3}%
\end{align}
In the c.m.f. we also have%
\begin{equation}
x_{q}+x_{\bar{q}}+z=1,\quad\vec{p}_{g}+\vec{p}_{q}+\vec{p}_{\bar{q}}=0.
\end{equation}

\subsection{Quark+gluon or antiquark+gluon in one jet}%
\label{SubSec:qg-and-qbarg-jets}

\begin{figure}[tbh]%
\hspace{0cm}\raisebox{4cm}{\includegraphics*[width=150mm
]%
{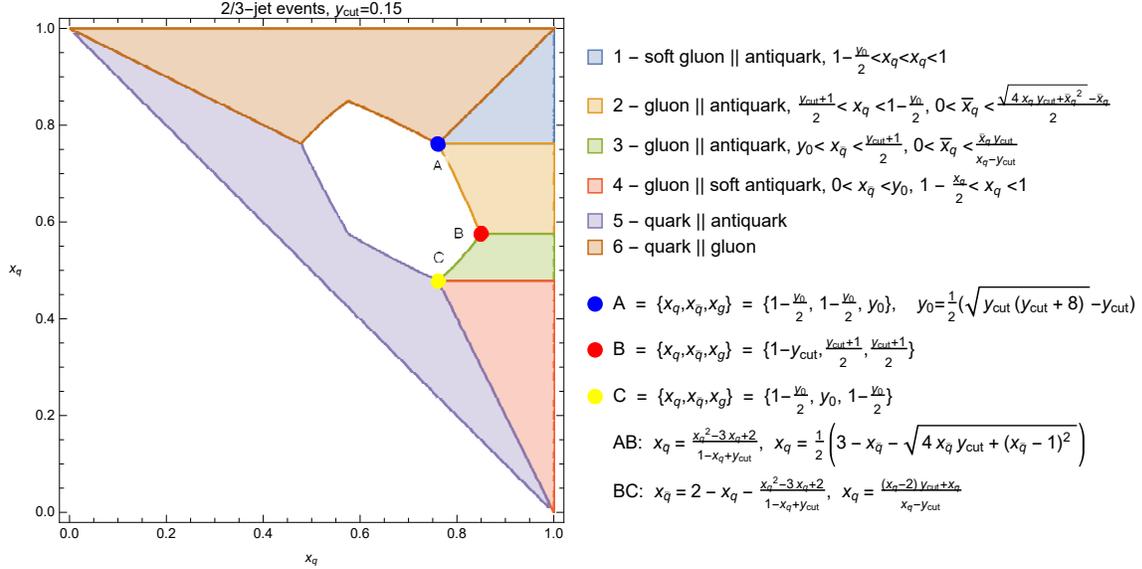}}
\vspace{-4cm}%
\caption{Dalitz plot for 2-3 jet separation in the exclusive $\ k_{t}$ algorithm
\cite{Brown:1991hx}. Regions 1--4 comprise the area of q- (\={q}g) dipole
configuration, i.e. collinear antiquark and gluon.
The dissection of the $q-(\bar{q}g)$ dipole area covering the curved polygon with the vertices $(1,1),\, A, \, B, \, C, \, (1,0)$ into regions is arbitrary. We found the tessellation depicted here convenient for integration.   
}%
\label{dalitzqqbar}%
\end{figure}
The integral over the area covered by regions 1--4 in figure \ref{dalitzqqbar} gives
the contribution of configurations where the antiquark and the gluon form one jet,
 jet i.e. when the gluon and the antiquark are almost collinear to each other.
The other jet is then formed by the quark.
So we have
\begin{align}
\vec{p}_{j}  &  =\vec{p}_{q},\quad x_{j}=x_{q},\quad\vec{p}_{\bar{j}}=\vec
{p}_{\bar{q}}+\vec{p}_{g},\quad x_{\bar{j}}=x_{\bar{q}}+z,\quad\vec{\Delta
}_{q}=\frac{x_{\bar{q}}\vec{p}_{g}-z\vec{p}_{\bar{q}}}{x_{\bar{q}}+z},\\
\quad\mathbf{\bar{x}}_{q}  &  \sim \vec{\Delta}_{q}^{\,2}\frac{x_{\bar{j}}^{2}%
}{z(x_{\bar{j}}-z)}\frac{x_{\bar{j}}x_{j}}{\vec{p}_{j}{}^{2}}<f(\mathbf{x}%
_{\bar{q}})\sim O(\sqrt{y_{cut}}),\quad\mathbf{\bar{x}}_{\bar{q}}=\frac
{z}{x_{\bar{j}}}+O(\sqrt{y_{cut}}),\\
\mathbf{\bar{x}}_{g}  &  =\frac{x_{\bar{q}}}{x_{\bar{j}}}+O(\sqrt{y_{cut}%
}),\quad M^2=\frac{\vec{p}_{j}^{\,\,2}}{x_{j}x_{\bar{j}}}+O(\sqrt{y_{cut}}),
\end{align}
which follows from (\ref{mass1}--\ref{mass3}). Here $f$ describes the inner
border of Regions 1--4 along the curve connecting the points (1,1), A, B, C, (1,0).

The cross section for $q\bar{q}g$ production has a contribution $d\sigma_{3}$
with 2 dipole operators, a contribution $d\sigma_{4}$ with a dipole operator
and a double dipole operator, and a contribution $d\sigma_{5}$ with 2 double
dipole operators (see (6.5--6.8) in ref.~\cite{Boussarie:2016ogo}),%
\begin{equation}
d\sigma_{(q\bar{q}g)}=d\sigma_{3}+d\sigma_{4}+d\sigma_{5}.
\end{equation}
Here $d\sigma_{3}$ describes final state interaction and contains collinear
and soft singularities while $d\sigma_{4}$ and $d\sigma_{5}$ are finite.
Collinear singularities lie at $\mathbf{x}_{q}=1$ and $\mathbf{x}_{\bar{q}}=1$
and the soft one is in the corner $\mathbf{x}_{q}=\mathbf{x}_{\bar{q}}=1$ in
figure \ref{dalitzqqbar}. In this paper we will work only with the singular
part of $d\sigma_{3}$, where (see (7.8) of ref.~\cite{Boussarie:2016ogo})%
\begin{equation}
d\sigma_{3}(x_{q},\vec{p}_{q})|_{\operatorname{col}}=d\sigma(x_{j},\vec{p}%
_{j})|_{Born}\,\alpha_{s}\frac{\Gamma(1-\epsilon)}{(4\pi)^{1+\epsilon}}%
\frac{N_{c}^{2}-1}{2N_{c}}n_{j},
\end{equation}
and the collinear factor $n_{\bar{j}}$\ (see (7.9) in \cite{Boussarie:2016ogo}%
) reads
\begin{equation}
n_{\bar{j}}=\frac{\mu^{-2\epsilon}}{\Gamma(1-\epsilon)\pi^{\frac{d}{2}}}%
\int_{\alpha}^{x_{\bar{j}}}\frac{dz}{z}\int_{\vec{\Delta}_{q}^{\,2}%
<f\left(\frac{x_{\bar{j}}-z}{x_{\bar{j}}}\right)\frac{z(x_{\bar{j}}-z)}{x_{\bar{j}}^{2}%
}\frac{\vec{p}_{j}{}^{2}}{x_{\bar{j}}x_{j}}}d^{d}\vec{\Delta}_{q}\frac
{dz^{2}+4x_{\bar{j}}\left(  x_{\bar{j}}-z\right)  }{x_{\bar{j}}^{2}\vec
{\Delta}_{q}^{2}}. \label{CollinearFactor}%
\end{equation}
Here we modified the integration area in $n_{\bar{j}}$ according to $k_{t}$
jet algorithm whereas in ref.~\cite{Boussarie:2016ogo} we used cone algorithm.
After integration we get%
\begin{align}
n_{\bar{j}}+n_{j}  &  =4\left[  \left(  \ln\left(  \frac{M^2}{\mu^{2}%
}\right)  +\frac{1}{\epsilon}\right)  \left(  \ln\left(  \frac{x_{j}x_{\bar
{j}}}{\alpha^{2}}\right)  -\frac{3}{2}\right)  -\frac{1}{2}\ln^{2}\left(
\frac{x_{j}x_{\bar{j}}}{\alpha^{2}}\right)  \right. \nonumber\\
&  -\left.  \frac{1}{2}\ln^{2}\left(  \frac{x_{j}}{x_{\bar{j}}}\right)
+w(y_{cut})\right]  ,
\end{align}
where
\begin{align}
&  w(y_{cut})=2Li_{2}\left(  -\frac{y_{0}}{2y_{cut}}\right)  -Li_{2}\left(
\frac{y_{0}^{2}}{4y_{cut}}\right)  +2Li_{2}\left(  \frac{1-y_{0}}{1-y_{cut}%
}\right)  +Li_{2}(y_{cut})\nonumber\\
&  +2Li_{2}(1-y_{0})-\ln^{2}\frac{y_{0}}{2}+\ln2\left(  2\ln(1-y_{0}%
)-\frac{5y_{cut}^{2}}{2}+7y_{cut}-\frac{9}{2}\right)  -\frac{2\pi^{2}}%
{3}\nonumber\\
&  +\ln y_{cut}\left(  \frac{y_{0}^{2}+2y_{0}-3y_{cut}^{2}+6y_{cut}-3}{2}%
+2\ln(1-y_{0})\right)  -\frac{y_{0}^{2}+2y_{0}+3}{2}\ln\frac{y_{0}}%
{2}\nonumber\\
&  +\frac{y_{cut}^{2}+2y_{cut}-y_{0}(y_{0}+2)}{2}\ln(y_{0}-y_{cut}%
)+\frac{3-y_{cut}^{2}-2y_{cut}}{2}\ln(1-y_{cut})\nonumber\\
&  +\frac{6y_{cut}^{3}+y_{cut}^{2}(y_{0}-20)+2y_{cut}\left(  y_{0}^{2}%
+7y_{0}+16\right)  +y_{0}\left(  y_{0}^{2}+10y_{0}+14\right)  }{4(2y_{cut}%
+y_{0})}\nonumber\\
&  +\ln^{2}\left(  \frac{1-y_{cut}}{1-y_{0}}\right)  +\frac{3}{2}%
(1-y_{cut})^{2}\ln(2y_{cut}+y_{0})+\frac{1}{2}\left(  y_{0}^{2}+2y_{0}%
-3\right)  \ln(1-y_{0}).
\end{align}
Here
\begin{equation}
y_{0}=\frac{\sqrt{y_{cut}(y_{cut}+8)}-y_{cut}}{2}.
\end{equation}
This result cancels soft and collinear singularities in the virtual part and
we get instead of (7.24) in ref.~\cite{Boussarie:2016ogo}%
\begin{align}
S_{R}  &  =n_{\bar{j}}+n_{j}+S_{V}+S_{V}^{\ast}\\
&  =4\left.  \left[  -\frac{1}{2}\ln^{2}\left(  \frac{x_{j}}{x_{\bar{j}}%
}\right)  +w(y_{cut})+3-\frac{\pi^{2}}{6}\right]  \right\vert _{y_{cut}%
=0.15}=4\left[  -\frac{1}{2}\ln^{2}\left(  \frac{x_{j}}{x_{\bar{j}}}\right)
+1.33\right]  .
\end{align}
In the small $y_{cut}$ approximation
\begin{equation}
S_{R}=4\left[  -\frac{1}{2}\ln^{2}\left(  \frac{x_{j}}{x_{\bar{j}}}\right)
-\frac{1}{2}\ln^{2}y_{cut}-\frac{3}{2}\ln y_{cut}-\frac{7\pi^{2}}{12}%
+\frac{13}{2}-\ln8\right]  +O(\sqrt{y_{cut}}).
\end{equation}
The remaining contributions of $d\sigma_{3},$ $d\sigma_{4},$ and $d\sigma_{5}$
are suppressed in $y_{cut}.$ Therefore the contribution of the soft and
collinear gluons to the cross section after cancellation of divergencies with
the virtual part reads%
\begin{equation}
d\sigma_{3}(x_{q},\vec{p}_{q})|_{\operatorname{col}}=d\sigma_{0}(x_{j},\vec
{p}_{j})\,\frac{\alpha_{s}}{4\pi}\frac{N_{c}^{2}-1}{2N_{c}}S_{R}%
+O(\sqrt{y_{cut}}).
\end{equation}
Nevertheless $O(\sqrt{y_{cut}})$ corrections for $y_{cut}=0.15$ are substantial,
e.g. the next (numerically largest) correction to $S_{R}$ reads
\begin{equation}
S_{R}=4\left[  -\frac{1}{2}\ln^{2}\left(  \frac{x_{j}}{x_{\bar{j}}}\right)
-0.29+4\sqrt{2y_{cut}}\right]  +O(y_{cut}),\quad4\sqrt{2y_{cut}}\simeq 2.19\,.
\end{equation}
It means that leading in $y_{cut}$ contribution is numerically of the same
order as $O(\sqrt{y_{cut}})$ corrections. But corrections of this order come
from all other contributions to the cross section, i.e. the remaining part of
$d\sigma_{3},$ $d\sigma_{4},$ and $d\sigma_{5}$ integrated over the whole
3--jet area (regions 1--6). Therefore the result for $S_{R}$ alone can not be
a good approximation. It has importance rather as a subtraction term for
future full numerical calculation.

Nevertheless using eq.~(\ref{dsigmaBorn}),%
\begin{eqnarray}
&&\left.\frac{d\sigma_{3}}{d\beta d\phi}\right|_{\operatorname{col}}   =\frac{2\alpha
^{2}Q_{q}^{2}N_{c}\sigma_{0}^{2}}{(2\pi)^{4}B_{G}(\hbar c)^{2}}\frac
{\alpha_{s}}{\pi}\frac{N_{c}^{2}-1}{2N_{c}}\nonumber\\
&&  \times\int_{(1-x_{P {\rm max}}\beta)\max\left(0.1s,\frac{Q_{\rm min}^2}{x_{P {\rm max}}\beta},\frac{Q_{\rm min}^2%
}{x_{P {\rm max}}\bar{\beta}}\right)}^{W_{\rm max}^{2}}\frac{dW^{2}}{s}\int_{\max\left(Q_{\rm min}^2,Q_{\rm min}^2%
\frac{\beta}{\bar{\beta}},0.1s-W^{2}\right)}^{\frac{x_{P {\rm max}}\beta}{1-x_{P {\rm max}}\beta}W^{2}%
}\frac{dQ^{2}}{yR_{0}^{4}Q^{6}}\nonumber\\
&&  \times\int_{x_{\min}}^{\bar{x}_{\min}}dx\left[  \bar{y}\frac{(\bar{\beta
}-\beta)^{2}\beta^{2}}{x\bar{x}}+\frac{[(1-2x)^{2}+1]}{2}\frac{\bar{\beta
}\beta^{3}}{(x\bar{x})^{2}}\left\{  \frac{1+\bar{y}^{2}}{2}-\frac{2\bar
{y}x\bar{x}}{1-2x\bar{x}}\cos(2\phi)\right\}  \right] \nonumber\\
&&  \times\left[  w(y_{cut})+3-\frac{\pi^{2}}{6}-\frac{1}{2}\ln^{2}\left(
\frac{\bar{x}}{x}\right)  \right]  .
\end{eqnarray}
Then the $x$ integral is doable analytically, see eq.~(\ref{dsigma/dbetaddfiAnalytic})%
\begin{eqnarray}
&&\hspace{-.2cm} \frac{d\sigma_{ep}}{d\beta d\phi}|_{\operatorname{col}}   =\frac{\alpha_{s}%
}{\pi}\frac{N_{c}^{2}-1}{2N_{c}}\left[  w(y_{cut})+3-\frac{\pi^{2}}{6}\right]
\frac{d\sigma_{ep}}{d\beta d\phi}|_{Born}\nonumber\\
&& \hspace{-.2cm} +\frac{2\alpha^{2}Q_{q}^{2}N_{c}\sigma_{0}^{2}}{(2\pi)^{4}B_{G}(\hbar
c)^{2}}\!\int_{(1-x_{P {\rm max}}\beta)\max\left(0.1s,\frac{Q_{\rm min}^2}{x_{P {\rm max}}\beta},\frac{Q_{\rm min}^2%
}{x_{P {\rm max}}\bar{\beta}}\right)}^{W_{\rm max}^{2}}\!\!\frac{dW^{2}}{s}\int_{\max\left(Q_{\rm min}^2,Q_{\rm min}^2%
\frac{\beta}{\bar{\beta}},0.1s-W^{2}\right)}^{\frac{x_{P {\rm max}}\beta}{1-x_{P {\rm max}}\beta}W^{2}%
}\frac{dQ^{2}}{yR_{0}^{4}Q^{6}}\nonumber\\
&& \hspace{-.2cm}  \times\frac{\alpha_{s}}{\pi}\frac{N_{c}^{2}-1}{2N_{c}}\left[  -\frac{1}%
{3}\ln^{3}\left(  \frac{\bar{x}_{\min}}{x_{\min}}\right)  \left(  \bar{y}%
(\bar{\beta}-\beta)^{2}\beta^{2}-2\bar{y}\bar{\beta}\beta^{3}\cos
(2\phi)\right)  \right. \nonumber\\
&& \hspace{-.2cm}  +\left.  \bar{\beta}\beta^{3}\frac{1+\bar{y}^{2}}{2}\frac{2(1-2x_{\min}%
\bar{x}_{\min})\ln\left(  \frac{\bar{x}_{\min}}{x_{\min}}\right)
-(1-2x_{\min})\left(  \ln^{2}\left(  \frac{\bar{x}_{\min}}{x_{\min}}\right)
+2\right)  }{x_{\min}\bar{x}_{\min}}\right]  . \label{collCorr}%
\end{eqnarray}
%

\begin{figure}[ptb]%
\centering
\raisebox{7.5cm}{\includegraphics[angle=-90,
width=130mm
]%
{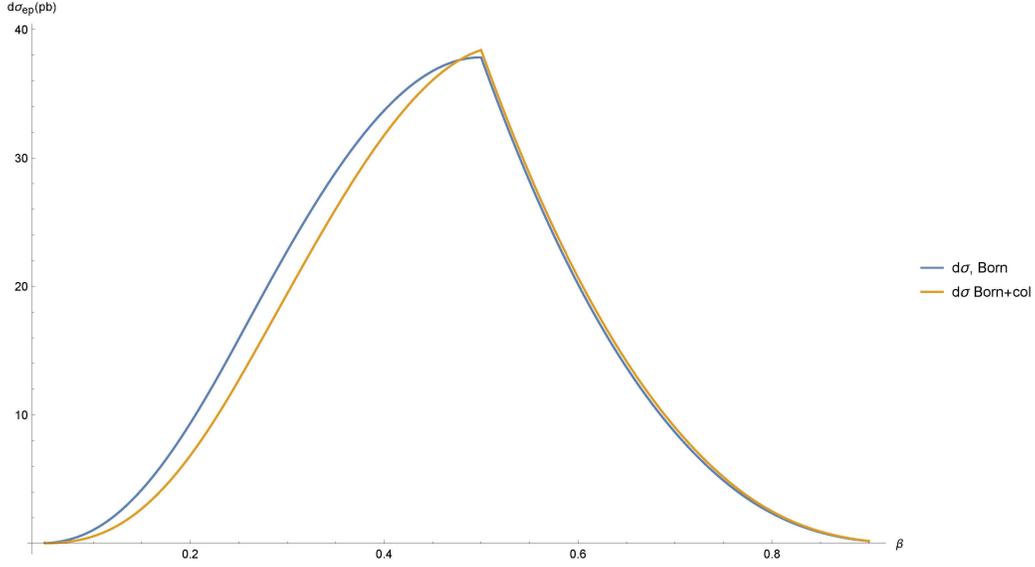}}%
\caption{Born (\ref{dsigma/dbetaddfiAnalytic}) and collinear correction
(\ref{collCorr}) to e-p cross section.}%
\label{collinearCorrection}%
\end{figure}
The result is given in figure \ref{collinearCorrection}. One may notice a sharp corner of the graph at $\beta=0.5$. It is related to the change of the functional dependence on $\beta$ in the limits of $Q$ and $W$ integrations of the cross section at $\beta=0.5$, which is a consequence of the HERA cuts.

\subsection{Quark+antiquark in one jet}
\label{SubSec:qqbar-jets}

\begin{figure}[ptb]%
\hspace{.5cm}\raisebox{2cm}{\includegraphics[angle=-90,
width=130mm
]%
{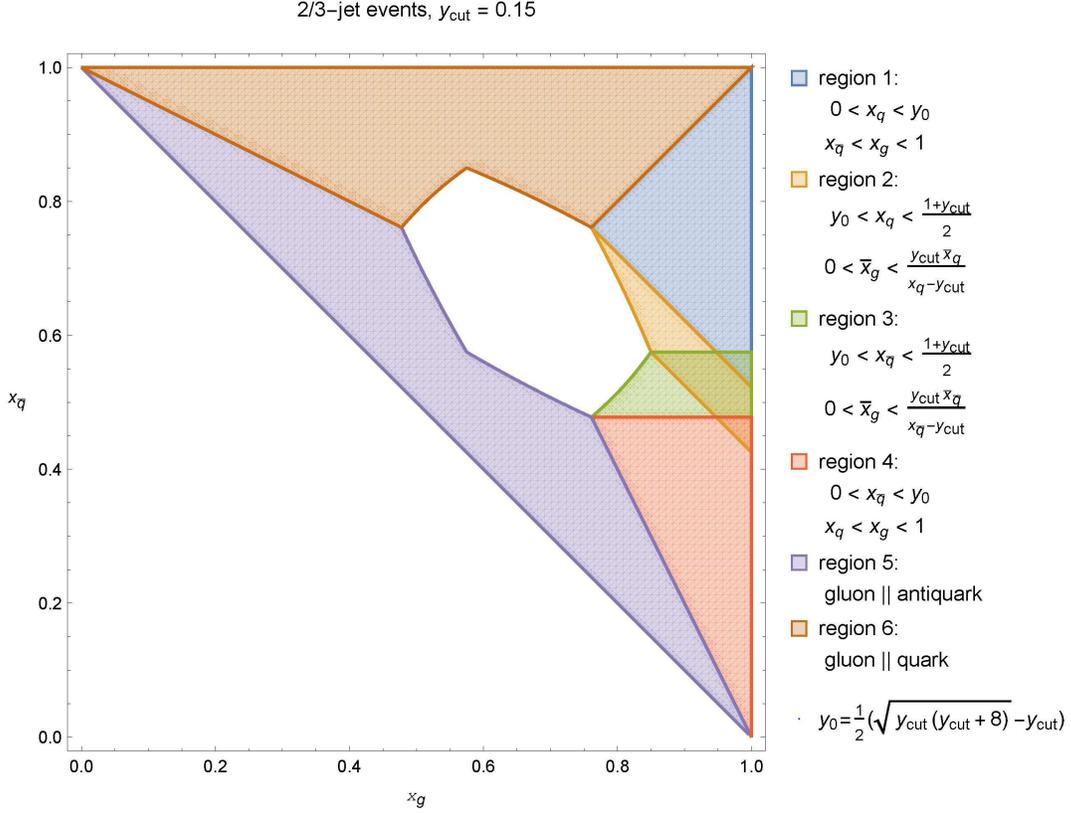}}
\vspace{.5cm}%
\caption{Dalitz plot for 2-3 jet separation in$\ k_{t}$ exclusive algorithm
\cite{Brown:1991hx}. Regions 1--4 comprise the area of gluon - (q\={q}) dipole
configuration.}%
\label{dalitzg}%
\end{figure}

The integral over the area covered by regions 1--4 in figure \ref{dalitzg}
is $\sim\sqrt{y_{cut}}.$ These regions cover the configurations with a collinear
quark-antiquark pair. However, this contribution may be enhanced in the large
produced mass $M$ limit thanks to the $t$-channel gluon in the impact factor.
In this picture collinear $q\bar{q}$ configurations cover regions 1--4, where
\begin{align}
\vec{p}_{j}  &  =\vec{p}_{g},\quad x_{j}=z,\quad\vec{p}_{\bar{j}}=\vec
{p}_{\bar{q}}+\vec{p}_{q},\quad x_{\bar{j}}=x_{\bar{q}}+x_{q},\quad\vec
{\Delta}_{g}=\frac{x_{\bar{q}}\vec{p}_{q}-x_{q}\vec{p}_{\bar{q}}}{x_{\bar{q}%
}+x_{q}},\\
\quad\mathbf{\bar{x}}_{g}  &  \sim \vec{\Delta}_{g}^{\,2}\frac{x_{\bar{j}}^{2}%
}{x_{q}(x_{\bar{j}}-x_{q})}\frac{x_{\bar{j}}x_{j}}{\vec{p}_{j}{}^{2}}%
<f(\mathbf{x}%
_{g}) \sim O(\sqrt{y_{cut}}),\quad\mathbf{\bar{x}}_{\bar{q}}=\frac{x_{q}}{x_{\bar{j}}%
}+O(\sqrt{y_{cut}}),\\
\mathbf{\bar{x}}_{q}  &  =\frac{x_{\bar{q}}}{x_{\bar{j}}}+O(\sqrt{y_{cut}%
}),\quad M^2=\frac{\vec{p}_{j}^{\,\,2}}{x_{j}x_{\bar{j}}}+O(\sqrt{y_{cut}}),
\end{align}
which follows from (\ref{mass1}--\ref{mass3}).

The cross section for $q\bar{q}g$ production has a contribution $d\sigma_{3}$
with 2 dipole operators, a contribution $d\sigma_{4}$ with a dipole operator
and a double dipole operator, and a contribution $d\sigma_{5}$ with 2 double
dipole operators (see (6.5--6.8) in ref.~\cite{Boussarie:2016ogo}),
see appendix \ref{append:normalization} for proper normalization,%
\begin{equation}
d\sigma_{(q\bar{q}g)}=d\sigma_{3}+d\sigma_{4}+d\sigma_{5}.\label{sigmaR}%
\end{equation}
Since the photon in the initial state can appear with different polarizations, the various cross sections are labeled as
\begin{equation}
d\sigma_{JI}=%
\begin{pmatrix}
d\sigma_{LL} & d\sigma_{LT}\\
d\sigma_{TL} & d\sigma_{TT}%
\end{pmatrix}
,\qquad d\sigma_{TL}=d\sigma_{LT}^{\ast}.
\end{equation}
The dipole $\times$ dipole contribution reads 
\begin{align}
d\sigma_{3JI} &  =\alpha_{s}\frac{N_{c}^{2}-1}{N_{c}}\frac{1}{2(2\pi)^{4}%
}\frac{\alpha_{\mathrm{em}}Q_{q}^{2}}{(2\pi)^{4}N_{c}}\frac{(p_{0}^{-})^{2}%
}{s^{2}x_{q}x_{\bar{q}}}\varepsilon_{I\alpha}\varepsilon_{J\beta}^{\ast}%
dx_{q}\,dx_{\bar{q}}\,d^{2}p_{q\bot}\,d^{2}p_{\bar{q}\bot}\frac{dzd^{2}%
p_{g\bot}}{z(2\pi)^{2}}\nonumber\\
&  \times\delta(1-x_{q}-x_{\bar{q}}-z)\int d^{2}p_{1\bot}d^{2}p_{2\bot}%
d^{2}p_{1\bot}^{\prime}d^{2}p_{2\bot}^{\prime}\delta(p_{q1\bot}+p_{\bar
{q}2\bot}+p_{g\bot})\nonumber\\
&  \times\delta(p_{11^{\prime}\bot}+p_{22^{\prime}\bot})\Phi_{3}^{\alpha
}(p_{1\bot},p_{2\bot})\Phi_{3}^{\beta\ast}(p_{1\bot}^{\prime},p_{2\bot
}^{\prime})\mathbf{F}\left(  \frac{p_{12\bot}}{2}\right)  \mathbf{F}^{\ast
}\left(  \frac{p_{1^{\prime}2^{\prime}\bot}}{2}\right)  \,.\label{dsigma3}%
\end{align}
The dipole $\times$ double dipole contribution reads
\begin{align}
d\sigma_{4JI} & \! =\!\frac{1}{2(2\pi)^{4}}\alpha_{s}\frac{\alpha_{\mathrm{em}%
}Q_{q}^{2}}{(2\pi)^{4}N_{c}}\frac{(p_{0}^{-})^{2}}{s^{2}x_{q}x_{\bar{q}}%
}(\varepsilon_{I\alpha}\varepsilon_{J\beta}^{\ast})dx_{q}dx_{\bar{q}}%
d^{2}p_{q\bot}d^{2}p_{\bar{q}\bot}\!\frac{dzd^{2}p_{g\bot}}{z(2\pi)^{2}}%
\delta(1-x_{q}-x_{\bar{q}}-z)\nonumber\\
&  \hspace{-1cm}\times\int d^{2}p_{1\bot}d^{2}p_{2\bot}d^{2}p_{1\bot}^{\prime
}d^{2}p_{2\bot}^{\prime}\frac{d^{2}p_{3\bot}d^{2}p_{3\bot}^{\prime}}{\left(
2\pi\right)  ^{2}}\delta(p_{q1\bot}+p_{\bar{q}2\bot}+p_{g3\bot})\delta
(p_{11^{\prime}\bot}+p_{22^{\prime}\bot}+p_{33^{\prime}\bot})\nonumber\\
&  \hspace{-1cm}\times\left[  \Phi_{3}^{\alpha}(p_{1\bot},p_{2\bot})\Phi
_{4}^{\beta\ast}(p_{1\bot}^{\prime},p_{2\bot}^{\prime},p_{3\bot}^{\prime
})\mathbf{F}\left(  \frac{p_{12\bot}}{2}\right)  \mathbf{\tilde{F}}^{\ast
}\left(  \frac{p_{1^{\prime}2^{\prime}\bot}}{2},p_{3\bot}^{\prime}\right)
\delta(p_{3\bot})\right.  \nonumber\\
&  \hspace{-1cm}+\left.  \Phi_{4}^{\alpha}(p_{1\bot},p_{2\bot},p_{3\bot}%
)\Phi_{3}^{\beta\ast}(\frac{p_{1^{\prime}2^{\prime}\bot}}{2})\mathbf{\tilde
{F}}\left(  \frac{p_{12\bot}}{2},p_{3\bot}\right)  \mathbf{F}^{\ast}\left(
\frac{p_{1^{\prime}2^{\prime}\bot}}{2}\right)  \delta(p_{3\bot}^{\prime
})\right]  .
\end{align}
The double dipole $\times$ double dipole contribution to the 3 jet cross
section reads
\begin{align}
\hspace{-1cm} &  d\sigma_{5JI}\!=\!\frac{1}{2(2\pi)^{4}}\alpha_{s}\frac
{\alpha_{\mathrm{em}}Q_{q}^{2}}{(2\pi)^{4}}\frac{(p_{0}^{-})^{2}}{s^{2}%
x_{q}x_{\bar{q}}}\frac{(\varepsilon_{I\alpha}\varepsilon_{J\beta}^{\ast}%
)}{N_{c}^{2}-1}dx_{q}dx_{\bar{q}}d^{2}p_{q\bot}d^{2}p_{\bar{q}\bot}%
\frac{dzd^{2}p_{g\bot}}{z(2\pi)^{2}}\delta(1-x_{q}-x_{\bar{q}}-z)\nonumber\\
&  \hspace{0cm}\times\int d^{2}p_{1\bot}d^{2}p_{2\bot}d^{2}p_{1\bot}^{\prime
}d^{2}p_{2\bot}^{\prime}\frac{d^{2}p_{3\bot}d^{2}p_{3\bot}^{\prime}}{\left(
2\pi\right)  ^{4}}\delta(p_{q1\bot}+p_{\bar{q}2\bot}+p_{g3\bot})\delta
(p_{11^{\prime}\bot}+p_{22^{\prime}\bot}+p_{33^{\prime}\bot})\nonumber\\
&  \hspace{0cm}\times\Phi_{4}^{\alpha}(p_{1\bot},p_{2\bot},p_{3\bot})\Phi
_{4}^{\beta\ast}(p_{1\bot}^{\prime},p_{2\bot}^{\prime},p_{3\bot}^{\prime
})\mathbf{\tilde{F}}_{p_{0\bot}p_{0\bot}^{\prime}}\left(  \frac{p_{12\bot}}%
{2},p_{3\bot}\right)  \mathbf{\tilde{F}}_{p_{0\bot}p_{0\bot}^{\prime}}^{\ast
}\left(  \frac{p_{1^{\prime}2^{\prime}\bot}}{2},p_{3\bot}^{\prime}\right)  .
\end{align} 
Here the impact factors are given in ref.~\cite{Boussarie:2016ogo} and  in Appendix~\ref{append:interference}, whereas the hadronic matrix
elements are given by eq.~(5.3) of ref.~\cite{Boussarie:2016ogo}. Changing variables%
\begin{equation}
\vec{p}_{q},\,\vec{p}_{\bar{q}},\,\vec{p}_{g},\,x_{q},\,\,z\,\rightarrow
\,\vec{p}=\vec{p}_{q}+\vec{p}_{\bar{q}}+\vec{p}_{g},\,\vec{p}_{j}%
,\,\vec{\Delta}_{g},\,x_{q},\,x_{j},
\end{equation}
one gets%
\begin{align}
d\sigma_{3JI} &  =\alpha_{s}\frac{N_{c}^{2}-1}{N_{c}^{2}}\frac{1}{2(2\pi)^{4}%
}\frac{\alpha_{\mathrm{em}}Q_{q}^{2}}{(2\pi)^{4}}\frac{(p_{0}^{-})^{2}}{s^{2}%
}(\varepsilon_{I\alpha}\varepsilon_{J\beta}^{\ast})\,\,\frac{dx_{q}d\vec
{p}_{j}d\vec{p}d\vec{\Delta}_{g}}{x_{q}(1-x_{q}-x_{j})}\frac{dx_{j}}%
{x_{j}(2\pi)^{2}}\nonumber\\
&  \times\int d^{2}p_{1\bot}d^{2}p_{2\bot}d^{2}p_{1\bot}^{\prime}d^{2}%
p_{2\bot}^{\prime}\delta(\vec{p}_{1}+\vec{p}_{2}-\vec{p})\delta(p_{11^{\prime
}\bot}+p_{22^{\prime}\bot})\nonumber\\
&  \times\Phi_{3}^{\alpha}(p_{1\bot},p_{2\bot})\Phi_{3}^{\beta\ast}(p_{1\bot
}^{\prime},p_{2\bot}^{\prime})\mathbf{F}_{p_{0\bot},p_{0\bot}-p_{\bot}}\left(
\frac{p_{12\bot}}{2}\right)  \mathbf{F}_{p_{0\bot},p_{0\bot}-p_{\bot}}^{\ast
}\left(  \frac{p_{1^{\prime}2^{\prime}\bot}}{2}\right)  \,,
\end{align}%
\begin{align}
d\sigma_{4JI} &  =\alpha_{s}\frac{1}{2(2\pi)^{4}}\frac{\alpha_{\mathrm{em}%
}Q_{q}^{2}}{(2\pi)^{4}}\frac{(p_{0}^{-})^{2}}{s^{2}}\frac{(\varepsilon
_{I\alpha}\varepsilon_{J\beta}^{\ast})}{N_{c}}\frac{dx_{q}}{x_{q}%
(1-x_{q}-x_{j})}\frac{dx_{j}}{x_{j}(2\pi)^{2}}d\vec{p}_{j}d\vec{p}d\vec
{\Delta}_{g}\nonumber\\
&  \hspace{-1cm}\times\int d^{2}p_{1\bot}d^{2}p_{2\bot}d^{2}p_{1\bot}^{\prime
}d^{2}p_{2\bot}^{\prime}\frac{d^{2}p_{3\bot}d^{2}p_{3\bot}^{\prime}}{\left(
2\pi\right)  ^{2}}\delta(\vec{p}_{1}+\vec{p}_{2}+\vec{p}_{3}-\vec{p}%
)\delta(p_{11^{\prime}\bot}+p_{22^{\prime}\bot}+p_{33^{\prime}\bot
})\nonumber\\
&  \hspace{-1cm}\times\left[  \Phi_{3}^{\alpha}(p_{1\bot},p_{2\bot})\Phi
_{4}^{\beta\ast}(p_{1\bot}^{\prime},p_{2\bot}^{\prime},p_{3\bot}^{\prime
})\mathbf{F}_{p_{0\bot},p_{0\bot}-p_{\bot}}\left(  \frac{p_{12\bot}}%
{2}\right)  \mathbf{\tilde{F}}_{p_{0\bot},p_{0\bot}-p_{\bot}}^{\ast}\left(
\frac{p_{1^{\prime}2^{\prime}\bot}}{2},p_{3\bot}^{\prime}\right)
\delta(p_{3\bot})\right.  \nonumber\\
&  \hspace{-1cm}+\left.  \Phi_{4}^{\alpha}(p_{1\bot},p_{2\bot},p_{3\bot}%
)\Phi_{3}^{\beta\ast}(\frac{p_{1^{\prime}2^{\prime}\bot}}{2})\mathbf{\tilde
{F}}_{p_{0\bot},p_{0\bot}-p_{\bot}}\left(  \frac{p_{12\bot}}{2},p_{3\bot
}\right)  \mathbf{F}_{p_{0\bot},p_{0\bot}-p_{\bot}}^{\ast}\left(
\frac{p_{1^{\prime}2^{\prime}\bot}}{2}\right)  \delta(p_{3\bot}^{\prime
})\right]  ,
\end{align}%
\begin{align}
&  d\sigma_{5JI}=\alpha_{s}\frac{1}{2(2\pi)^{4}}\frac{\alpha_{\mathrm{em}%
}Q_{q}^{2}}{(2\pi)^{4}}\frac{(p_{0}^{-})^{2}}{s^{2}}\frac{(\varepsilon
_{I\alpha}\varepsilon_{J\beta}^{\ast})}{N_{c}^{2}-1}\frac{dx_{q}}%
{x_{q}(1-x_{j}-x_{q})}\frac{dx_{j}}{x_{j}(2\pi)^{2}}d\vec{p}_{j}d\vec{p}%
d\vec{\Delta}_{g}\nonumber\\
&  \times\int d^{2}p_{1\bot}d^{2}p_{2\bot}d^{2}p_{1\bot}^{\prime}d^{2}%
p_{2\bot}^{\prime}\frac{d^{2}p_{3\bot}d^{2}p_{3\bot}^{\prime}}{\left(
2\pi\right)  ^{4}}\delta(\vec{p}_{1}+\vec{p}_{2}+\vec{p}_{3}-\vec{p}%
)\delta(p_{11^{\prime}\bot}+p_{22^{\prime}\bot}+p_{33^{\prime}\bot
})\nonumber\\
&  \times\Phi_{4}^{\alpha}(p_{1\bot},p_{2\bot},p_{3\bot})\Phi_{4}^{\beta\ast
}(p_{1\bot}^{\prime},p_{2\bot}^{\prime},p_{3\bot}^{\prime})\mathbf{\tilde{F}%
}_{p_{0\bot},p_{0\bot}-p_{\bot}}\!\!\left(  \frac{p_{12\bot}}{2},p_{3\bot}\!\right)\!
\mathbf{\tilde{F}}_{p_{0\bot},p_{0\bot}-p_{\bot}}^{\ast}\!\!\left(  \frac
{p_{1^{\prime}2^{\prime}\bot}}{2},p_{3\bot}^{\prime}\!\right) \! .
\end{align}
The hadronic matrix\ elements can be written as (see (5.2--5.8) in ref.~\cite{Boussarie:2016ogo})%
\begin{align}
& \  2\pi\delta(p_{00^{\prime}}^{-})\mathbf{\tilde{F}}_{p_{0\bot}p_{0\bot
}^{\prime}}(\frac{p_{12\bot}}{2},p_{3\bot})\nonumber\\
& \! =2\pi\delta(p_{00^{\prime}}^{-})N_{c}(2\pi)^{2}\left[  \delta(\vec{p}%
_{2})\mathbf{F}_{p_{0\bot}p_{0\bot}^{\prime}}(\frac{p_{13\bot}}{2}%
)+\delta(\vec{p}_{1})\mathbf{F}_{p_{0\bot}p_{0\bot}^{\prime}}(\frac{p_{32\bot
}}{2})-\delta(\vec{p}_{3})\mathbf{F}_{p_{0\bot}p_{0\bot}^{\prime}}%
(\frac{p_{12\bot}}{2})\right]  \nonumber\\
&  +\!\!\int \! d\vec{x} \, d\vec{y} \, e^{-i[(\frac{\vec{x}-\vec{y}}{2}\cdot\vec{p}%
_{12})+(\frac{\vec{x}+\vec{y}}{2}\cdot\vec{p}_{3})]} \nonumber\\
& \times \langle P^{\prime}%
(p_{0}^{\prime})|T\left[  (tr(U_{\frac{x-y}{2}}U_{\frac{x+y}{2}}^{\dag}%
)-N_{c})(tr(U_{\frac{x+y}{2}}U_{\frac{y-x}{2}}^{\dag})-N_{c})\right]
|P(p_{0})\rangle.
\end{align}
As a first approximation one may neglect the nonlinear term. Then we have%
\begin{align}
&  \mathbf{\tilde{F}}_{p_{0\bot}p_{0\bot}^{\prime}}(\frac{p_{12\bot}}%
{2},p_{3\bot})\nonumber\\
&  \simeq N_{c}(2\pi)^{2}\left[  \delta(\vec{p}_{2})\mathbf{F}_{p_{0\bot
}p_{0\bot}^{\prime}}(\frac{p_{13\bot}}{2})+\delta(\vec{p}_{1})\mathbf{F}%
_{p_{0\bot}p_{0\bot}^{\prime}}(\frac{p_{32\bot}}{2})-\delta(\vec{p}%
_{3})\mathbf{F}_{p_{0\bot}p_{0\bot}^{\prime}}(\frac{p_{12\bot}}{2})\right]  \\
&  =N_{c}(2\pi)^{2}\left[  \delta(\vec{p}_{2})\mathbf{F}(\frac{p_{13\bot}}%
{2})+\delta(\vec{p}_{1})\mathbf{F}(\frac{p_{32\bot}}{2})-\delta(\vec{p}%
_{3})\mathbf{F}(\frac{p_{12\bot}}{2})\right]  e^{-\frac{B_{G}}{2}\vec
{p}^{\,\,2}}\\
&  =-4\pi(2\pi)^{2}N_{c}^{2}\sigma_{0}\left[  \delta(\vec{p}_{2})e^{-R_{0}%
^{2}\vec{p}_{1}^{\,\,2}}\frac{\partial}{\partial p_{1}^{2}}+\delta(\vec{p}%
_{1})e^{-R_{0}^{2}\vec{p}_{3}^{\,\,2}}\frac{\partial}{\partial p_{3}^{2}%
}-\delta(\vec{p}_{3})e^{-R_{0}^{2}\vec{p}_{2}^{\,\,2}}\frac{\partial}{\partial
p_{2}^{2}}\right]  e^{-\frac{B_{G}}{2}\vec{p}^{\,\,2}}.
\end{align}
Intrinsically this assumes large $N_{c}$ approximation so that we will neglect
$1$ in $N_{c}^{2}-1.$ Integrating w.r.t. $\vec{p}$ via
\begin{equation}
\int e^{-B_{G}\vec{p}^{\,\,2}}d\vec{p}=\frac{\pi}{B_{G}}%
\end{equation}
and substituting
\begin{equation}
dx_{j}d\vec{p}_{j}=dx_{j}d\phi\frac{d\vec{p}_{j}^{\,\,2}}{2}\rightarrow
d\phi\frac{d\beta}{2\beta^{2}}Q^{2}\int_{x_{\min}}^{x_{\max}}x_{j}x_{\bar{j}%
}dx_{j},\label{xj-integral}%
\end{equation}
one gets%
\begin{align}
\frac{d\sigma_{3JI}}{d\beta d\phi} &  =\frac{1}{2}N_{c}^{2}\sigma_{0}^{2}%
\frac{\alpha_{s}\alpha_{\mathrm{em}}Q_{q}^{2}}{(2\pi)^{7}B_{G}}\frac
{(p_{0}^{-})^{2}Q^{2}}{s^{2}\beta^{2}}(\varepsilon_{I\alpha}\varepsilon
_{J\beta}^{\ast})\int_{x_{\min}}^{x_{\max}}x_{\bar{j}}dx_{j}\,\int%
_{\text{Regions 1-4}}\frac{dx_{q}d\vec{\Delta}_{g}}{x_{q}(1-x_{q}-x_{j}%
)}\nonumber\\
&  \times\int d^{2}p_{1\bot}d^{2}p_{1\bot}^{\prime}e^{-R_{0}^{2}\vec{p}%
_{1}^{\,\,2}}\frac{\partial}{\partial p_{1}^{2}}\Phi_{3}^{\alpha}(p_{1\bot
},-p_{1\bot})e^{-R_{0}^{2}\vec{p}_{1}^{\prime\,\,2}}\frac{\partial}{\partial
p_{1}^{\prime2}}\Phi_{3}^{\beta\ast}(p_{1\bot}^{\prime},-p_{1\bot}^{\prime
})\,,\label{dsigma3dp}%
\end{align}%
\begin{align}
& \frac{d\sigma_{4JI}}{d\beta d\phi}   =\frac{1}{2}N_{c}^{2}\sigma_{0}^{2}%
\frac{\alpha_{s}\alpha_{\mathrm{em}}Q_{q}^{2}}{(2\pi)^{7}B_{G}}\frac
{(p_{0}^{-})^{2}Q^{2}}{s^{2}\beta^{2}}(\varepsilon_{I\alpha}\varepsilon
_{J\beta}^{\ast})\int_{x_{\min}}^{x_{\max}}x_{\bar{j}}dx_{j}\int%
_{\text{Regions 1-4}}\frac{dx_{q}d\vec{\Delta}_{g}}{x_{q}(1-x_{q}-x_{j}%
)}\nonumber\\
&  \times\int\left[  \left(  \delta(\vec{p}_{2}^{\prime})e^{-R_{0}^{2}\vec
{p}_{1}^{\prime\,\,2}}\frac{\partial}{\partial p_{1}^{\prime2}}+\delta(\vec
{p}_{1}^{\prime})e^{-R_{0}^{2}\vec{p}_{3}^{\prime\,\,2}}\frac{\partial
}{\partial p_{3}^{\prime2}}-\delta(\vec{p}_{3}^{\prime})e^{-R_{0}^{2}\vec
{p}_{2}^{\prime\,\,2}}\frac{\partial}{\partial p_{2}^{\prime2}}\right)
\Phi_{4}^{\beta\ast}(p_{1\bot}^{\prime},p_{2\bot}^{\prime},p_{3\bot}^{\prime
})\right.  \nonumber\\
&  \times e^{-R_{0}^{2}\vec{p}_{1}^{\,\,2}}\frac{\partial}{\partial p_{1}^{2}%
}\Phi_{3}^{\alpha}(p_{1\bot},p_{2\bot})\delta(p_{3\bot})+e^{-R_{0}^{2}\vec
{p}_{1}^{\prime\,\,2}}\frac{\partial}{\partial p_{1}^{\prime2}}\Phi_{3}%
^{\beta\ast}(\frac{p_{1^{\prime}2^{\prime}\bot}}{2})\delta(p_{3\bot}^{\prime
})\nonumber\\
&  \times\left.  \left(  \delta(\vec{p}_{2})e^{-R_{0}^{2}\vec{p}_{1}^{\,\,2}%
}\frac{\partial}{\partial p_{1}^{2}}+\delta(\vec{p}_{1})e^{-R_{0}^{2}\vec
{p}_{3}^{\,\,2}}\frac{\partial}{\partial p_{3}^{2}}-\delta(\vec{p}%
_{3})e^{-R_{0}^{2}\vec{p}_{2}^{\,\,2}}\frac{\partial}{\partial p_{2}^{2}%
}\right)  \Phi_{4}^{\alpha}(p_{1\bot},p_{2\bot},p_{3\bot})\right]  \nonumber\\
&  \times d^{2}p_{1\bot}d^{2}p_{2\bot}d^{2}p_{3\bot}d^{2}p_{1\bot}^{\prime
}d^{2}p_{2\bot}^{\prime}d^{2}p_{3\bot}^{\prime}\delta(\vec{p}_{1}+\vec{p}%
_{2}+\vec{p}_{3})\delta(p_{11^{\prime}\bot}+p_{22^{\prime}\bot}+p_{33^{\prime
}\bot}),
\end{align}%
\begin{align}
& \frac{d\sigma_{5JI}}{d\beta d\phi}=\frac{1}{2}   N_{c}^{2}\sigma_{0}^{2}%
\frac{\alpha_{s}\alpha_{\mathrm{em}}Q_{q}^{2}}{(2\pi)^{7}B_{G}}\frac
{(p_{0}^{-})^{2}Q^{2}}{s^{2}\beta^{2}}(\varepsilon_{I\alpha}\varepsilon
_{J\beta}^{\ast})\int_{x_{\min}}^{x_{\max}}x_{\bar{j}}dx_{j}\int%
_{\text{Regions 1-4}}\frac{dx_{q}d\vec{\Delta}_{g}}{x_{q}(1-x_{j}-x_{q}%
)}\nonumber\\
\times &  \int\left[  \delta(\vec{p}_{2})e^{-R_{0}^{2}\vec{p}_{1}^{\,\,2}%
}\frac{\partial}{\partial p_{1}^{2}}+\delta(\vec{p}_{1})e^{-R_{0}^{2}\vec
{p}_{3}^{\,\,2}}\frac{\partial}{\partial p_{3}^{2}}-\delta(\vec{p}%
_{3})e^{-R_{0}^{2}\vec{p}_{2}^{\,\,2}}\frac{\partial}{\partial p_{2}^{2}%
}\right]  \Phi_{4}^{\alpha}(p_{1\bot},p_{2\bot},p_{3\bot})\,\nonumber\\
\times &  \left[  \delta(\vec{p}_{2}^{\prime})e^{-R_{0}^{2}\vec{p}_{1}%
^{\prime\,\,2}}\frac{\partial}{\partial p_{1}^{\prime2}}+\delta(\vec{p}%
_{1}^{\prime})e^{-R_{0}^{2}\vec{p}_{3}^{\prime\,\,2}}\frac{\partial}{\partial
p_{3}^{\prime2}}-\delta(\vec{p}_{3}^{\prime})e^{-R_{0}^{2}\vec{p}_{2}%
^{\prime\,\,2}}\frac{\partial}{\partial p_{2}^{\prime2}}\right]  \Phi
_{4}^{\beta}(p_{1\bot}^{\prime},p_{2\bot}^{\prime},p_{3\bot}^{\prime})^{\ast
}\nonumber\\
&  \times d^{2}p_{1\bot}d^{2}p_{2\bot}d^{2}p_{3\bot}d^{2}p_{1\bot}^{\prime
}d^{2}p_{2\bot}^{\prime}d^{2}p_{3\bot}^{\prime}\delta(\vec{p}_{1}+\vec{p}%
_{2}+\vec{p}_{3})\delta(p_{11^{\prime}\bot}+p_{22^{\prime}\bot}+p_{33^{\prime
}\bot}).\label{dsigma5dp}%
\end{align}
First, one has to integrate these expressions over the area covered by Regions
1--4 in the Dalitz plot (fig.~\ref{dalitzg}). In terms of the plot variables
$\mathbf{x}$ the integral reads%
\begin{align}
\int_{\text{Regions 1-4}}d\mathbf{x}_{g}d\mathbf{x}_{q} &  =\int%
_{\text{Regions 1-4}}d\mathbf{x}_{g}d\mathbf{x}_{\bar{q}}\nonumber\\
&  =\left[  \int_{1-y_{0}}^{1}d\mathbf{\bar{x}}_{q}\int_{0}^{\frac
{\mathbf{x}_{q}}{2}}d\mathbf{\bar{x}}_{g}+\int_{\frac{1-y_{cut}}{2}}^{1-y_{0}%
}d\mathbf{\bar{x}}_{q}\int_{0}^{\frac{y_{cut}\mathbf{\bar{x}}_{q}}%
{\mathbf{x}_{q}-y_{cut}}}d\mathbf{\bar{x}}_{g}+(\mathbf{x}_{q}\leftrightarrow
\mathbf{x}_{\bar{q}})\right]  \nonumber\\
&  -\int_{\frac{1-y_{cut}}{2}}^{\frac{1+y_{cut}}{2}}d\mathbf{\bar{x}}_{q}%
\int_{0}^{\frac{1+y_{cut}}{2}-\mathbf{\bar{x}}_{q}}d\mathbf{\bar{x}}_{g}.
\end{align}
Since the impact factor is symmetric w.r.t. $q\leftrightarrow\bar{q}$
interchange, one can rewrite the latter expression as%
\begin{align}
\int_{\text{Regions 1-4}}d\mathbf{x}_{g}d\mathbf{x}_{q} &  =2\int_{1-y_{0}%
}^{1}d\mathbf{\bar{x}}_{q}\int_{0}^{\frac{\mathbf{x}_{q}}{2}}d\mathbf{\bar{x}%
}_{g}+2\int_{\frac{1-y_{cut}}{2}}^{1-y_{0}}d\mathbf{\bar{x}}_{q}\int%
_{0}^{\frac{y_{cut}\mathbf{\bar{x}}_{q}}{\mathbf{x}_{q}-y_{cut}}}%
d\mathbf{\bar{x}}_{g}\nonumber\\
&  -\int_{\frac{1-y_{cut}}{2}}^{\frac{1+y_{cut}}{2}}d\mathbf{\bar{x}}_{q}%
\int_{0}^{\frac{1+y_{cut}}{2}-\mathbf{\bar{x}}_{q}}d\mathbf{\bar{x}}_{g}.
\end{align}
The impact factors are not singular as $\Delta_{g}\equiv|\vec{\Delta}_{g}|\rightarrow0$ and
$\vec{\Delta}_{g}^{\,2}\sim\bar{\mathbf{x}}_{g}.$ Therefore to get the leading in
$\sqrt{y_{cut}}$ contribution, one can put $\Delta_{g}=0$ in them and
integrate w.r.t. $\Delta_{g}$%
\begin{align}
\int_{\text{Regions 1-4}}dx_{q}d\vec{\Delta}_{g} &  =\frac{\vec{p}_{j}{}^{2}%
}{x_{\bar{j}}x_{j}}\pi x_{\bar{j}}\int_{\text{Regions 1-4}}d\mathbf{x}%
_{g}d\mathbf{x}_{\bar{q}}\frac{x_{q}(x_{\bar{j}}-x_{q})}{x_{\bar{j}}^{2}%
}\nonumber\\
&  =2\pi\frac{\vec{p}_{j}{}^{2}}{x_{\bar{j}}x_{j}}\int_{x_{\bar{j}}(1-y_{0}%
)}^{x_{\bar{j}}}dx_{q}\frac{1}{2}\frac{x_{\bar{j}}-x_{q}}{x_{\bar{j}}}%
\frac{x_{q}(x_{\bar{j}}-x_{q})}{x_{\bar{j}}^{2}}\nonumber\\
&  +2\pi\frac{\vec{p}_{j}{}^{2}}{x_{\bar{j}}x_{j}}\int_{x_{\bar{j}}%
\frac{1-y_{cut}}{2}}^{x_{\bar{j}}(1-y_{0})}dx_{q}y_{cut}\frac{1-\frac
{x_{\bar{j}}-x_{q}}{x_{\bar{j}}}}{\frac{x_{\bar{j}}-x_{q}}{x_{\bar{j}}%
}-y_{cut}}\frac{x_{q}(x_{\bar{j}}-x_{q})}{x_{\bar{j}}^{2}}+O(y_{cut}%
).\label{q||qbarRegion}%
\end{align}
Next, we will work in the small $Q_{s}$ approximation as we did for the LO
impact factor (see eqs.~(\ref{smallQs1}--\ref{loTifSmallQsGBW})). It means that
after integrating out delta-functions and calculating derivatives in
eqs.~(\ref{dsigma3dp}--\ref{dsigma5dp}), one takes the angular integrals of the
remaining $t$-channel momenta ($\vec{p}_{1}^{(\prime)},\vec{p}_{2}^{(\prime
)},$ or $\vec{p}_{3}^{(\prime)}$) and neglects their absolute values
everywhere except in the exponents. Then the exponential integrals are calculated
straightforwardly giving $\int_{0}^{+\infty}dp^{2}e^{-R_{0}^{2}p^{\,2}}%
=\frac{1}{R_{0}^{2}}.$ As a result one has the following cross sections%
\begin{align}
\hspace{-1cm}\frac{d\sigma_{LL}}{d\beta d\phi} &  =\frac{1}{2}N_{c}^{2}%
\sigma_{0}^{2}\frac{\alpha_{s}\alpha_{\mathrm{em}}Q_{q}^{2}}{(2\pi)^{4}B_{G}%
}\frac{\sqrt{2y_{cut}}}{R_{0}^{4}Q^{4}}a+O\left(  y_{cut}\right)
,\label{resultLL}\\
\frac{d\sigma_{TT}^{ij}}{d\beta d\phi} &  =\frac{1}{2}N_{c}^{2}\sigma_{0}%
^{2}\frac{\alpha_{s}\alpha_{\mathrm{em}}Q_{q}^{2}}{(2\pi)^{4}B_{G}}\frac
{\sqrt{2y_{cut}}}{R_{0}^{4}Q^{4}}(cg_{\bot}^{ij}+be^{(x)i}e^{(x)j})+O\left(
y_{cut}\right)  ,\label{resultTT}\\
a &  =a_{3}+a_{4}+a_{5},\quad b=b_{3}+b_{4}+b_{5},\quad c=c_{3}+c_{4}%
+c_{5}.\label{abc}%
\end{align}
Here $a_{i},b_{i},c_{i},$ $i=3,4,5$ are the contributions coming from
eqs.~(\ref{dsigma3dp}--\ref{dsigma5dp}) correspondingly.

We demonstrate this procedure on the example of the longitudinal photon
contribution to $\sigma_{5}.$ The impact factor for longitudinal photon $\times$
longitudinal photon was calculated in ref.~\cite{Boussarie:2016ogo} (B.1). It
reads
\begin{eqnarray}
&&\frac{(p_{0}^{-})^{2}}{s^{2}}\varepsilon_{L\alpha}\Phi_{4}^{\alpha}(p_{1\bot
},p_{2\bot},p_{3\bot})\,\varepsilon_{L\beta}^{\ast}\Phi_{4}^{\beta}(p_{1\bot
}^{\prime},p_{2\bot}^{\prime},p_{3\bot}^{\prime})^{\ast} \\
&&=4Q^{2}\frac
{(x_{q}^{2}+(x_{q}+z)^{2})x_{\bar{q}}^{2}}{x_{q}x_{\bar{q}}z^{2}}\vec{V}%
_{q}(p_{1},p_{3})\vec{V}_{q}(p_{1}^{\prime},p_{3}^{\prime}) \nonumber \\
&&-4Q^{2}\frac{(x_{\bar{q}}+z)x_{q}+(x_{q}+z)x_{\bar{q}}}{z^{2}}\vec{V}%
_{q}(p_{1},p_{3})\vec{V}_{\bar{q}}(p_{2}^{\prime},p_{3}^{\prime})+(x_{q}%
,p_{1},p_{1}^{\prime},V_{q}\leftrightarrow x_{\bar{q}},p_{2},p_{2}^{\prime
},V_{\bar{q}})\,,\nonumber
\end{eqnarray}
where
\begin{equation}
\vec{V}_{q}(p_{1},p_{3})=\frac{x_{q}\vec{p}_{g3}{}-z\vec{p}_{q1}{}}{\left(
x_{q}+z\right)  \left(  \frac{(\vec{p}_{g3}+\vec{p}_{q1})^{2}}{x_{\bar{q}}%
}+\frac{\vec{p_{g3}}{}^{2}}{z}+\frac{\vec{p_{q1}}{}^{2}}{x_{q}}+Q^{2}\right)
\left(  \frac{(\vec{p}_{g3}+\vec{p}_{q1})^{2}}{x_{\bar{q}}\left(
x_{q}+z\right)  }+Q^{2}\right)  }.
\end{equation}
As was outlined above, using small $Q_{s}$ and small $y_{cut}$ approximations,
one can take $t$-channel integrals%
\begin{align}
&  \int\left[  \delta(\vec{p}_{2})e^{-R_{0}^{2}\vec{p}_{1}^{\,\,2}}%
\frac{\partial}{\partial p_{1}^{2}}+\delta(\vec{p}_{1})e^{-R_{0}^{2}\vec
{p}_{3}^{\,\,2}}\frac{\partial}{\partial p_{3}^{2}}-\delta(\vec{p}%
_{3})e^{-R_{0}^{2}\vec{p}_{2}^{\,\,2}}\frac{\partial}{\partial p_{2}^{2}%
}\right] \nonumber\\
&  \times\delta(\vec{p}_{1}+\vec{p}_{2}+\vec{p}_{3})\vec{V}_{q}(p_{1}%
,p_{3})d^{2}p_{1\bot}d^{2}p_{2\bot}d^{2}p_{3\bot}\nonumber\\
=  &  \int\left[  \delta(\vec{p}_{2})e^{-R_{0}^{2}\vec{p}_{1}^{\,\,2}}%
\frac{\partial}{\partial p_{1}^{2}}+\delta(\vec{p}_{1})e^{-R_{0}^{2}\vec
{p}_{3}^{\,\,2}}\frac{\partial}{\partial p_{3}^{2}}-\delta(\vec{p}%
_{3})e^{-R_{0}^{2}\vec{p}_{2}^{\,\,2}}\frac{\partial}{\partial p_{2}^{2}%
}\right] \nonumber\\
&  \times\delta(\vec{p}_{1}+\vec{p}_{2}+\vec{p}_{3})\left.  \vec{V}_{\bar{q}%
}(p_{2},p_{3})d^{2}p_{1\bot}d^{2}p_{2\bot}d^{2}p_{3\bot}\right\vert
_{x_{q}\rightarrow x_{\bar{q}}=1-x_{j}-x_{q}}\nonumber\\
\simeq &  \frac{\pi\beta^{3}\vec{p}_{j}}{R_{0}^{2}Q^{6}}\left(  \frac
{(\beta+1)x_{q}+2x_{j}(x_{\bar{j}}-x_{q})}{\left(  x_{j}\left(  x_{\bar{j}%
}-x_{q}\right)  +\beta x_{q}\right)  {}^{2}}-\frac{4}{x_{\bar{j}}x_{j}%
}\right)  .
\end{align}
Then one integrates over regions 1--4 via eq.~(\ref{q||qbarRegion}) and w.r.t.
$x_{j}$ according to eq.~(\ref{xj-integral}). 
Keeping only the leading contribution $y_{cut}$,
one gets
\begin{equation}
\hspace{-1cm}a_{5}=8\beta^{2}\bar{\beta}^{2}\ln\frac{\bar{x}_{\min}}{x_{\min}%
}. \label{dsigma5llrezult}%
\end{equation}

\begin{figure}[ptb]%
\hspace{0.2cm}
\raisebox{5.5cm}{\includegraphics[angle=-90,
width=50mm
]%
{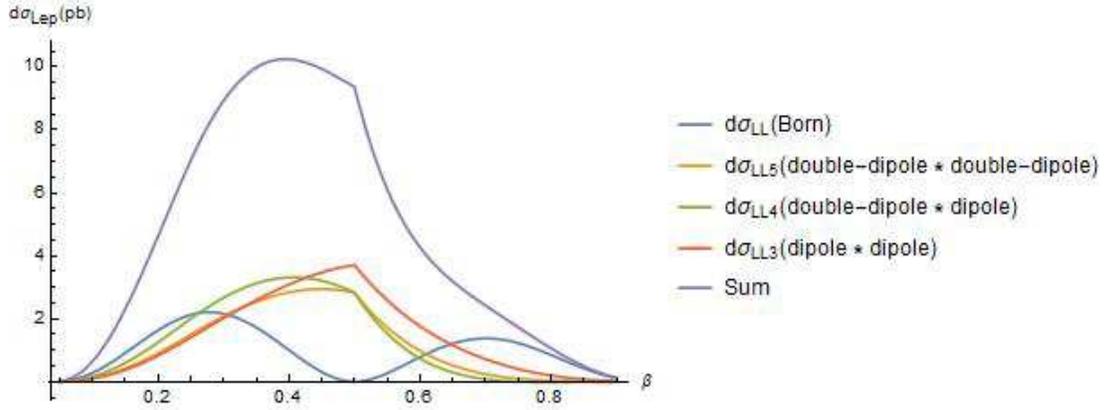}}%
\caption{$ep\rightarrow ep+2jets$ cross-section in the case of a longitudinal
photon.\ Born and gluon dipole contributions.}%
\label{sigmal}%
\end{figure}

\begin{figure}[ptb]%
\hspace{0.4cm}
\raisebox{5.2cm}{\includegraphics[angle=-90,
width=19mm
]%
{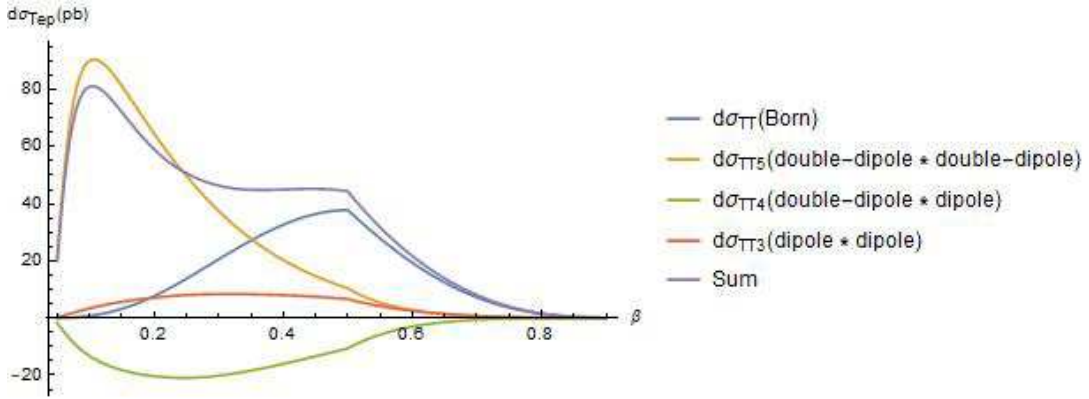}}%
\caption{$ep\rightarrow ep+2jets$ cross-section in the case of a transverse
photon.\ Born and gluon dipole contributions.}%
\label{sigmat1}%
\end{figure}

\begin{figure}[ptb]%
\hspace{.4cm}
\raisebox{6.8cm}{\includegraphics[angle=-90,
width=40mm
]%
{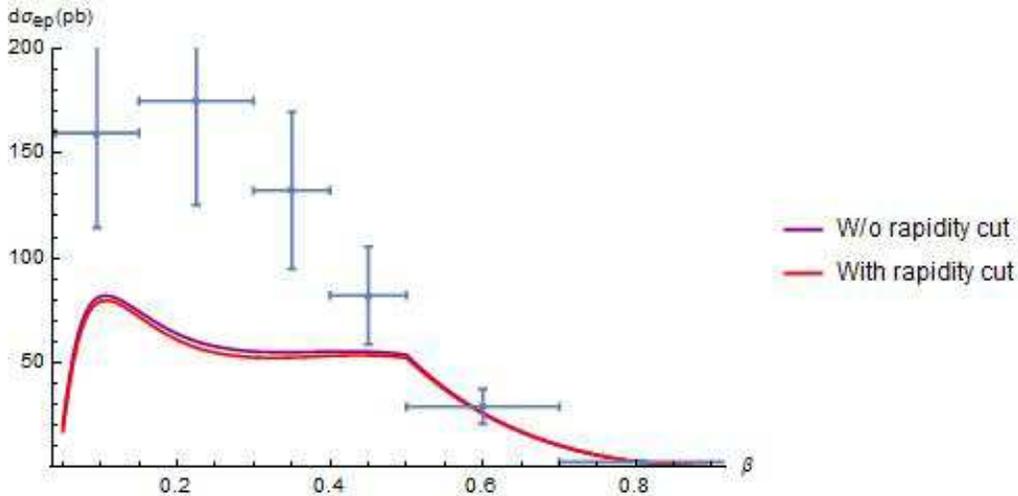}}%
\caption{Born and total gluon dipole contributions to cross section vs
experimental data from \cite{Abramowicz:2015vnu}. Rapidity cut is defined in
(\ref{rapiditycut1}--\ref{rapiditycut2}).}%
\label{SigmaGluonTotal1}%
\end{figure}

The product of the transverse photon $\times$ transverse photon impact factor\\ $\Phi_{4}%
^{i}(p_{1\bot},p_{2\bot},p_{3\bot})\Phi_{4}^{j}(p_{1\bot}^{\prime},p_{2\bot
}^{\prime},p_{3\bot}^{\prime})^{\ast}$ was calculated in
ref.~\cite{Boussarie:2016ogo}, see eq.~(B.16). The integration in this case is similar 
to the previous case, albeit with more cumbersome expressions. Therefore we do not
present the intermediate results giving only the final answer%
\begin{align}
b_{5} &  =2\bar{\beta}^{2}\left(  4\beta^{2}-1\right)  \ln\frac{\bar{x}_{\min
}}{x_{\min}},\\
c_{5} &  =\frac{\bar{\beta}}{4\beta}\left(  8\bar{\beta}\beta^{3}%
-2\beta-1\right)  \frac{1-2x_{\min}}{x_{\min}\bar{x}_{\min}}+\bar{\beta
}\left(  4\bar{\beta}\beta^{2}-1\right)  \ln\frac{\bar{x}_{\min}}{x_{\min}}.
\end{align}
The longitudinal photon $\times$ longitudinal photon impact factor $\Phi_{3}%
^{+}(p_{1\bot},p_{2\bot})\Phi_{3}^{+}(p_{1\bot}^{\prime},p_{2\bot}^{\prime
})^{\ast}$ was calculated in eqs.~(B.2--4) and the transverse one in eqs.~(B.17--19)
in ref.~\cite{Boussarie:2016ogo}. They lead to%
\begin{align}
a_{3} &  =\beta^{2}\left(  (4\beta(2\beta-3)+7)\ln\frac{\bar{x}_{\min}%
}{x_{\min}}-(1-2x_{\min})\right)  ,\\
b_{3} &  =\bar{\beta}\left(  \left(  \bar{\beta}+4\beta^{2}-8\beta^{3}\right)
\ln\frac{\bar{x}_{\min}}{x_{\min}}+(1+\beta)(1-2x_{\min})\right)  ,\\
c_{3} &  =\frac{\bar{\beta}}{4\beta}\left(  \beta^{2}\frac{4\beta-8\beta
^{2}-3}{x_{\min}\bar{x}_{\min}}-1\right)  (1-2x_{\min})+\frac{\bar{\beta}}%
{2}\left(  4\beta^{2}+1\right)  \left(  1-2\beta\right)  \ln\frac{\bar
{x}_{\min}}{x_{\min}}.
\end{align}
The remaining cross section $d\sigma_{4JI}$ contains $\Phi_{4}(p_{1\bot
},p_{2\bot},p_{3\bot})\Phi_{3}(p_{1\bot}^{\prime},p_{2\bot}^{\prime})^{\ast}.$
We present these convolutions in the Appendix~\ref{append:interference}. Integrating them according to
the guidelines discussed above we get
\begin{align}
a_{4} &  =4\beta^{2}\bar{\beta}(3-4\beta)\ln\frac{\bar{x}_{\min}}{x_{\min}},\\
b_{4} &  =2\bar{\beta}\left(  6\beta^{2}-8\beta^{3}-1\right)  \ln\frac{\bar
{x}_{\min}}{x_{\min}},\\
c_{4} &  =\frac{\bar{\beta}\beta}{2}\left(  1+6\beta-8\beta^{2}\right)
\frac{1-2x_{\min}}{x_{\min}\bar{x}_{\min}}+\frac{\bar{\beta}}{2\beta}\left(
2\beta^{2}(6\beta-8\beta^{2}-1)+1\right)  \ln\frac{\bar{x}_{\min}}{x_{\min}}.
\end{align}
Now one can recall the relation between the $\gamma^{\ast}P$ and $eP$ cross
sections (\ref{sigmagammaPtoeP}) and write, see eqs.~(\ref{resultLL}--\ref{abc}),
\begin{align}
&  \left.  \frac{d\sigma_{ep}}{d\beta d\phi}\right\vert
_{\substack{gluon\\dipole}}=\frac{\alpha^{2}\alpha_{s}Q_{q}^{2}}{B_{G}(\hbar
c)^{2}}\frac{\sqrt{2y_{cut}}}{(2\pi)^{5}}N_{c}^{2}\sigma_{0}^{2}%
\int_{(1-x_{P {\rm max}}\beta)\max(0.1s,\frac{Q_{\rm min}^2}{x_{P {\rm max}}\beta},\frac{Q_{\rm min}^2}%
{x_{P {\rm max}}\bar{\beta}})}^{W_{\rm max}^{2}}\frac{dW^{2}}{s}\nonumber\\
&  \times\int_{\max(Q_{\rm min}^2,Q_{\rm min}^2\frac{\beta}{\bar{\beta}},0.1s-W^{2})}%
^{\frac{x_{P {\rm max}}\beta}{1-x_{P {\rm max}}\beta}W^{2}}\frac{dQ^{2}}{yR_{0}^{4}Q^{6}}\left[
2\bar{y}a+\frac{1+\bar{y}^{2}}{2}(b-2c)+\bar{y}b\cos(2\phi)\right]
.\label{resultall}%
\end{align}
To get the distribution in $\beta$ one has to integrate this equation w.r.t.
$\phi$ from $0$ to $\pi$ because jets are treated as identical. The results
are in figures \ref{sigmal}, \ref{sigmat1}, \ref{SigmaGluonTotal1}.%
As one can see, the interference term$\ d\sigma_{4T}$ is negative, which
significantly diminishes the leading power asymptotics of $d\sigma_{5T}.$ In
addition, the large $N_{c}$ approximation decreases $d\sigma_{5T}$ for $\sim10\%$
since we expand $\frac{N_{c}^{4}}{N_{c}^{2}-1}\simeq N_{c}^{2}.$ On the other
hand the rapidity cut (\ref{rapiditycut1}--\ref{rapiditycut2}) dependence is very low.

\section{Conclusion}
\label{Sec:Conclusion}

This paper discussed the exclusive diffractive dijet electroproduction with
HERA\ selection cuts \cite{Abramowicz:2015vnu}. We started from the analytic
formulas from ref.~\cite{Boussarie:2016ogo} for fully differential Born cross
section and its real correction with dipole $\times$ dipole and double dipole
$\times$ double dipole configurations. In addition, in appendix \ref{append:interference} we
calculated the remaining interference real production impact factor with
dipole $\times$ double dipole configuration. We used the GBW parametrization
for the dipole matrix element between the proton states and the large $N_{c}$
approximation for the double dipole matrix elements. We constructed the differential
$ep\rightarrow ep+2jets$ cross section in $\beta=\frac{Q^{2}%
}{Q^{2}+M_{2jets}^{2}}$ and in the angle $\phi$ between the leptonic and hadronic
planes with HERA acceptance. We argued that HERA selection rules
\cite{Abramowicz:2015vnu} suppress the aligned jet contribution indicative of
saturation to the Born cross section. These cuts allowed us to neglect the
$t$-channel momentum in the Born impact factor and integrate the $\gamma p$
cross section analytically. The result is in eq.~(\ref{dsigma/dbetaddfiAnalytic}).

Next, we cancelled the singularities from soft and collinear gluons between
real and virtual corrections in the collinear approximation by integrating the
singular contributions over the $q-(\bar{q}g)$ and $\bar{q}-(qg)$ areas in the
Dalitz plot of fig.~\ref{dalitzqqbar}\ within the $k_{t}$ jet algorithm. As the Born
cross section, the resulting correction was analytically integrated in the
small $Q_{s}$ approximation in ref.~(\ref{collCorr}). It gives $\sim10\%$ of the
Born result.

Finally, we integrated all real corrections in the small $Q_{s}$ and small
$y_{cut}$ approximations over the the $g-(q\bar{q})$ area in the Dalitz plot
of fig.~\ref{dalitzg}\ within the $k_{t}$ jet algorithm. This configuration gives the
dominant contribution in the small $\beta$ region thanks to Regge enhancement
because of the diagram with $t$-channel gluon at large $s=M_{2jet}^{2}.$ The
results for this gluon dipole configuration are in refs.~(\ref{resultLL}--\ref{abc})
and refs.~(\ref{dsigma5llrezult}--\ref{resultall}). Results for Born and gluon
dipole in the small $Q_{s}$ approximation together give about $\frac{1}{2}$ of
the measured cross section.

We noted that firstly, the small $Q_{s}$ approximation works for Born,
collinearly enhanced radiative corrections to $q\bar{q}$ dipole configuration,
and generic gluon dipole configuration since the HERA cuts $Q,$ $M_{2jets}>5$
GeV and $p_{\bot\min}>2$ GeV effectively restrict jets with very small
longitudinal momentum fraction $x.$ It means that the typical hard scale in
the impact factor is of order of $M_{2jets}^{2},$ $Q^{2},$ $M_{2jets}^{2}x,$
$Q^{2}x,$ $p_{\bot\min}^{2}$ and multiplication with $x$ here can not make it
smaller than $Q_{s}^{2}.$ So we can expand the impact factor in $Q_{s}.$
However for Born, the region $x<Q_{s}^{2}/\max(Q^{2},M_{2jets}^{2})$ is the
aligned jet region indicative of saturation. 

Secondly, this approximation fails for other corrections to $q\bar{q}$ dipole
configuration since $Q_{s}$ may be the largest scale in the impact factor in
them. It also fails for gluon dipole configuration when the $q\bar{q}$ pair
forming one of the jets is in the aligned jet configuration itself since the
longitudinal momentum fraction of $q$ or $\bar{q}$ may be the small parameter
making the impact factor scales smaller than $Q_{s}$.

Thirdly, we nevertheless calculated the gluon dipole contribution in the small
$Q_{s}$ approximation neglecting that it may be incorrect in the
aforementioned corners of the phase space. Therefore comparison of our answer
to the full numerical result will show how important these contributions are. This is left for future studies.

Finally, we noted that the corrections in $y_{cut}$ may be significant since
the real expansion parameter is $\sqrt{y_{cut}}=\sqrt{0.15}\simeq0.39.$
Therefore the $O(\sqrt{y_{cut}})$ corrections to $q\bar{q}$ dipole
configuration which we did not calculate may give sizable corrections. However
we expect these corrections as well as the nonsingular virtual corrections to
be peaked at moderate $\beta$ as the Born term.

\acknowledgments

We thank G.~Gach, K.~Golec-Biernat, D.~Yu.~Ivanov, C.~Marquet, S.~Munier, B.~Schenke for many useful discussions.
A.~V.~G. acknowledges support of RFBR grant 19-02-00690. This work was supported
in part by the Ministry of Education and Science of the Russian Federation. He thanks the GDR QCD for support.
The work of R.~B. is supported by the U.S. Department of Energy,
Office of Nuclear Physics,
under Contracts No.  DE-SC0012704 and by an LDRD grant from Brookhaven Science
Associates. L.~S. is supported by the grant 2017/26/M/ST2/01074 of the National Science Center in Poland. He thanks the French LABEX P2IO, the French GDR QCD and the LPT for support. Part of the calculations were done on the supercomputer of Novosibirsk State University.

\appendix

\section{Scaling of the aligned versus symmetric jet contributions}
\label{append:Scaling}

As was discussed in the text, $F_{T}^{D(3)}$ is a higher twist correction when
compared to ref.~(\ref{GBstructureFunctions}) as it has an extra power of
$Q^{2}R_{0}^{2}\gg1$ in its denominator. The origin of this suppression lies in
the fact that the dominant contribution to the transverse cross section comes
from the aligned jet configuration, i.e. the region of $x\lesssim\frac{1}%
{\max(Q^{2},M^2)R_{0}^{2}}\ll1$. One can see it from eq.~(\ref{loTifExactGBW}):%
\begin{align}
A &  =\frac{\sigma_{0}}{\left\vert \vec{p}_{q\bar{q}}\right\vert }\int%
_{0}^{+\infty}dp^{2}e^{-R_{0}^{2}\vec{p}^{\,\,2}}\frac{\partial}{\partial
p^{2}}\frac{x\bar{x}Q^{2}+p^{2}-(\frac{\vec{p}_{q\bar{q}}{}}{2})^{2}}%
{\sqrt{(x\bar{x}Q^{2}+p^{2}+(\frac{\vec{p}_{q\bar{q}}{}}{2})^{2})^{2}%
-4p^{2}(\frac{\vec{p}_{q\bar{q}}{}}{2})^{2}}}\\
&  \sim\int_{0}^{1}da\frac{\sigma_{0}M Q^{2}R_{0}^{2}x^{\frac{3}{2}}}{\left[
(x(M^2+Q^{2})R_{0}^{2}+a)^{2}-4M^2R_{0}^{2}ax\right]  ^{\frac{3}{2}}}\\
&  =\frac{\sigma_{0}}{m\sqrt{x}}\left(  \frac{M^2-Q^{2}}{M^2+Q^{2}}%
+\frac{1+xR_{0}^{2}(Q^{2}-M^2)}{\sqrt{1+2xR_{0}^{2}(Q^{2}-M^2)+x^{2}%
R_{0}^{4}(Q^{2}+M^2)^{2}}}\right)  ,
\end{align}
where we approximated $e^{-R_{0}^{2}\vec{p}^{\,\,2}}\simeq\theta(R_{0}%
^{-2}-\vec{p}^{\,\,2}).$ Then we get the known behavior of
eq.~(\ref{GBstructureFunctions})%
\begin{equation}
x_{P}F_{T}^{D(3)}|_{Q^{2}\gg M^2}\sim\frac{Q^{4}}{\beta B_{G}}\int%
_{0}^{\frac{1}{Q^{2}R_{0}^{2}}}A^{2}xdx\sim\frac{Q^{4}}{\beta^{2}B_{G}}%
\frac{M^2\sigma_{0}^{2}}{Q^{6}R_{0}^{2}}=\frac{\bar{\beta}\sigma_{0}^{2}%
}{B_{G}R_{0}^{2}},
\end{equation}%
\begin{equation}
x_{P}F_{T}^{D(3)}|_{M^2\gg Q^{2}}\sim\frac{Q^{4}}{\beta B_{G}}\int%
_{0}^{\frac{1}{M^2R_{0}^{2}}}A^{2}xdx\sim\frac{Q^{4}}{\beta B_{G}}%
\frac{\sigma_{0}^{2}}{M^{4}R_{0}^{2}}=\frac{\beta\sigma_{0}^{2}}{B_{G}%
R_{0}^{2}}.
\end{equation}
It is easier to observe in the coordinate space (following ref.~\cite{GolecBiernat:2001hab}%
), where eqs.~(\ref{sigmaloL}--\ref{sigmaloT}) can be cast into
\begin{equation}
\frac{d\sigma_{T,L}^{\gamma p}}{dM^2}\!=\!\frac{3\alpha}{32\pi^{2}B_{G}}\!\sum
e_{f}^{2}\!\!\int\! dx \! \left\{
\begin{array}
[c]{c}%
\! \! (x^{2}+\bar{x}^{2})x^{2}\bar{x}^{2}Q^{2}\left(  \int rdrJ_{1}(\sqrt{x\bar{x}%
}Mr)K_{1}(\sqrt{x\bar{x}}Qr)\hat{\sigma}(r)\right)  ^{2}\\
4x^{3}\bar{x}^{3}Q^{2}\left(  \int rdrJ_{0}(\sqrt{x\bar{x}}Mr)K_{0}%
(\sqrt{x\bar{x}}Qr)\hat{\sigma}(r)\right)  ^{2}%
\end{array}
\!\right\}\!  ,
\end{equation}%
\begin{equation}
x_{P}F_{T,L}^{D(3)}=\frac{Q^{4}}{4\pi^{2}\alpha\beta}\frac{d\sigma
_{T,L}^{\gamma p}}{dM^2},\quad\hat{\sigma}(r)=\sigma_{0}(1-e^{-\frac{r^{2}%
}{4R_{0}^{2}}}).
\end{equation}
In the large $\beta$ region $Q^{2}R_{0}^{2}\gg1,$ $Q^{2}\gg\frac{1}{R_{0}^{2}%
}\gg M^2$ the longitudinal cross section reads%

\begin{align}
&  \int rdrJ_{0}(\sqrt{x\bar{x}}Mr)K_{0}(\sqrt{x\bar{x}}Qr)\hat{\sigma}%
(r)\sim\left(  \int_{0}^{\left(  \frac{2}{Q}\right)  ^{2}}+\int_{\frac{2}{Q}%
}^{\frac{1}{x\bar{x}Q^{2}}}\right)  dr^{2}\hat{\sigma}(r)\nonumber\\
&  \sim\int_{0}^{\left(  \frac{2}{Q}\right)  ^{2}}\frac{r^{2}}{R_{0}^{2}%
}dr^{2}+\theta\left(x>\frac{1}{Q^{2}R_{0}^{2}}\right)\int_{\left(  \frac{2}{Q}\right)
^{2}}^{\frac{1}{xQ^{2}}}\frac{r^{2}}{R_{0}^{2}}dr^{2}\nonumber\\
&  +\theta\left(x<\frac{1}{Q^{2}R_{0}^{2}}\right)\int_{\left(  \frac{2}{Q}\right)  ^{2}%
}^{R_{0}^{2}}\frac{r^{2}}{R_{0}^{2}}dr^{2}+\theta(x<\frac{1}{Q^{2}R_{0}^{2}%
})\int_{R_{0}^{2}}^{\frac{1}{xQ^{2}}}dr^{2}\\
&  \sim\frac{1}{Q^{4}R_{0}^{2}}+\theta\left(x>\frac{1}{Q^{2}R_{0}^{2}}\right)\frac
{1}{x^{2}Q^{4}R_{0}^{2}}+\theta\left(x<\frac{1}{Q^{2}R_{0}^{2}}\right)\frac{1}{xQ^{2}},
\end{align}
where neglecting logarithms
\begin{equation}
J_{0}(\sqrt{x\bar{x}}Mr)\sim1,\quad K_{0}(\sqrt{x\bar{x}}Qr)\sim\theta
(1-\sqrt{x\bar{x}}Qr).
\end{equation}%
\begin{eqnarray}
\frac{d\sigma_{L}^{\gamma p}}{dM^2}   &\sim& \int_{0}^{1}(x\bar{x}%
)^{3}dx\left[\left(  \frac{1}{Q^{4}R_{0}^{2}}\right)  ^{2}+\theta\left(x>\frac{1}%
{Q^{2}R_{0}^{2}}\right)\left(  \frac{1}{xQ^{4}R_{0}^{2}}\right)  ^{2}\right. \nonumber\\
&+&\theta\left(x    >\frac{1}{Q^{2}R_{0}^{2}}\right)\left(  \frac{1}{x^{2}Q^{4}R_{0}^{2}%
}\right)  ^{2}+\theta\left(x<\frac{1}{Q^{2}R_{0}^{2}}\right)\frac{1}{xQ^{2}}\frac
{1}{Q^{4}R_{0}^{2}}\nonumber\\
&+& \left. \theta\left(x    <\frac{1}{Q^{2}R_{0}^{2}}\right)\left(  \frac{1}{xQ^{2}}\right)
^{2}\right],
\end{eqnarray}%
\begin{equation}
\frac{d\sigma_{L}^{\gamma p}}{dM^2}\sim\frac{1}{Q^{8}R_{0}^{4}}+\frac
{1}{Q^{8}R_{0}^{4}}\ln\frac{1}{Q^{2}R_{0}^{2}}+\left(  \frac{1}{Q^{2}R_{0}%
^{2}}\right)  ^{3}\frac{1}{Q^{6}R_{0}^{2}}+\frac{1}{Q^{8}R_{0}^{4}} \, .
\end{equation}
So at $\beta\sim1,$ $x_{P}F_{L}^{D(3)}\sim\frac{1}{Q^{4}R_{0}^{4}}$ and and
this dominant contribution comes from the whole region in $x.$

The transverse cross section in the large $\beta$ region $Q^{2}R_{0}^{2}\gg1,$
$Q^{2}\gg\frac{1}{R_{0}^{2}}\gg M^2$ reads
\begin{align}
&  \int rdrJ_{1}(\sqrt{x\bar{x}}Mr)K_{1}(\sqrt{x\bar{x}}Qr)\hat{\sigma}%
(r)\sim\frac{\sqrt{x\bar{x}}Mr}{\sqrt{x\bar{x}}Qr}\left(  \int_{0}^{\frac
{2}{Q}}+\int_{\frac{2}{Q}}^{\frac{1}{x\bar{x}Q^{2}}}\right)  dr^{2}\hat
{\sigma}(r)\nonumber\\
&  \sim\frac{M}{Q}\left(  \int_{0}^{\left(  \frac{2}{Q}\right)  ^{2}}%
\frac{r^{2}}{R_{0}^{2}}dr^{2}+\theta\left(x>\frac{1}{Q^{2}R_{0}^{2}}\right)\int_{\left(
\frac{2}{Q}\right)  ^{2}}^{\frac{1}{x\bar{x}Q^{2}}}\frac{r^{2}}{R_{0}^{2}%
}dr^{2} \right.\nonumber\\
& \left.+\theta\left(x<\frac{1}{Q^{2}R_{0}^{2}}\right)\left(  \int_{\frac{1}{Q^{2}}%
}^{R_{0}^{2}}\frac{r^{2}}{R_{0}^{2}}+\int_{R_{0}^{2}}^{\frac{1}{x\bar{x}Q^{2}%
}}\right)  dr^{2}\right) \nonumber\\
&  \sim\frac{M}{Q}\left[  \frac{1}{Q^{4}R_{0}^{2}}+\theta\left(x>\frac{1}%
{Q^{2}R_{0}^{2}}\right)\frac{1}{x^{2}Q^{4}R_{0}^{2}}+\theta\left(x<\frac{1}{Q^{2}%
R_{0}^{2}}\right)\frac{1}{xQ^{2}}\right]  ,
\end{align}
where
\begin{equation}
J_{1}(\sqrt{x\bar{x}}Mr)\sim\sqrt{x\bar{x}}Mr,\quad K_{1}(\sqrt{x\bar{x}%
}Qr)\sim\frac{\theta(1-\sqrt{x\bar{x}}Qr)}{\sqrt{x\bar{x}}Qr}.
\end{equation}%
\begin{eqnarray}
&&\frac{d\sigma_{T}^{\gamma p}}{dM^2}    \sim\frac{M^2}{Q^{2}}\int_{0}%
^{1}(x\bar{x})^{2}dx\left[\left(  \frac{1}{Q^{4}R_{0}^{2}}\right)  ^{2}%
+\theta\left(x>\frac{1}{Q^{2}R_{0}^{2}}\right)\frac{1}{x^{2}}\left(  \frac{1}{Q^{4}%
R_{0}^{2}}\right)  ^{2}\right.\nonumber\\
&&+\theta\left(x    >\frac{1}{Q^{2}R_{0}^{2}}\right)\left(  \frac{1}{x^{2}Q^{4}R_{0}^{2}%
}\right)  ^{2}+\theta\left(x<\frac{1}{Q^{2}R_{0}^{2}}\right)\left(  \frac{1}{xQ^{2}%
}\right)  \frac{1}{Q^{4}R_{0}^{2}}\nonumber\\
&&\left.+\theta\left(x    <\frac{1}{Q^{2}R_{0}^{2}}\right)\left(  \frac{1}{xQ^{2}}\right)
^{2}\right],
\end{eqnarray}
\begin{eqnarray}
\frac{d\sigma_{T}^{\gamma p}}{dM^2}\sim\frac{M^2}{Q^{2}}\left[  \frac
{1}{Q^{8}R_{0}^{4}}+\frac{Q^{2}R_{0}^{2}}{Q^{8}R_{0}^{4}}+\left(  \frac
{1}{Q^{2}R_{0}^{2}}\right)  ^{2}\frac{1}{Q^{6}R_{0}^{2}}+\frac{1}{Q^{4}}%
\frac{1}{Q^{2}R_{0}^{2}}\right]  .
\end{eqnarray}
So at $\beta\sim1,$ $x_{P}F_{T}^{D(3)}\sim\frac{1}{Q^{4}R_{0}^{2}}$ and this
dominant contribution comes from $x<\frac{1}{Q^{2}R_{0}^{2}},$ i.e.
aligned jets.

In the small $\beta$ region $Q^{2}R_{0}^{2}\gg1,$ $M^2R_{0}^{2}\gg1,$
$M^2\gg Q^{2}\gg\frac{1}{R_{0}^{2}}$ for the longitudinal cross section we have%

\begin{align}
&  \int rdrJ_{0}(\sqrt{x\bar{x}}Mr)K_{0}(\sqrt{x\bar{x}}Qr)\hat{\sigma}%
(r)\sim\left(  \int_{0}^{\frac{2}{M}}+\int_{\frac{2}{M}}^{\frac{1}{x\bar
{x}M^2}}\right)  dr^{2}\hat{\sigma}(r)\nonumber\\
&  \sim\!\int_{0}^{\left(  \frac{2}{M}\right)  ^{2}}\!\!\!\frac{r^{2}}{R_{0}^{2}%
}dr^{2}+\theta\left(x>\frac{1}{M^2R_{0}^{2}}\right)\!\!\int_{\left(  \frac{2}{M}\right)
^{2}}^{\frac{1}{xM^2}}\!\frac{r^{2}}{R_{0}^{2}}dr^{2}+\theta\left(x<\frac{1}%
{M^2R_{0}^{2}}\right)\!\!\left(  \int_{\frac{1}{M^2}}^{R_{0}^{2}}\frac{r^{2}}%
{R_{0}^{2}}+\int_{R_{0}^{2}}^{\frac{1}{xM^2}}\right)  dr^{2}\nonumber\\
&  \sim\frac{1}{M^{4}R_{0}^{2}}+\theta\left(x>\frac{1}{M^2R_{0}^{2}}\right)\frac
{1}{x^{2}M^{4}R_{0}^{2}}+\theta\left(x<\frac{1}{M^2R_{0}^{2}}\right)\frac{1}{xM^2},
\end{align}
where again neglecting logarithms
\begin{equation}
J_{0}(\sqrt{x\bar{x}}Mr)\sim\theta(1-\sqrt{x\bar{x}}Mr),\quad K_{0}%
(\sqrt{x\bar{x}}Qr)\sim1.
\end{equation}%
\begin{eqnarray}
&&\frac{d\sigma_{L}^{\gamma p}}{dM^2}    \sim\int_{0}^{1}(x\bar{x}%
)^{3}dx\left[\left(  \frac{1}{M^{4}R_{0}^{2}}\right)  ^{2}+\theta(x>\frac{1}%
{M^2R_{0}^{2}})\left(  \frac{1}{xM^{4}R_{0}^{2}}\right)  ^{2}\right.\nonumber\\
&&+\theta(x    >\frac{1}{M^2R_{0}^{2}})\left(  \frac{1}{x^{2}M^{4}R_{0}^{2}%
}\right)  ^{2}+\theta(x<\frac{1}{M^2R_{0}^{2}})\left(  \frac{1}{xM^2%
}\right)  \frac{1}{M^{4}R_{0}^{2}}\nonumber\\
&&\left.+\theta(x    <\frac{1}{M^2R_{0}^{2}})\left(  \frac{1}{xM^2}\right)
^{2}\right],
\end{eqnarray}%
i.e.
\begin{equation}
\frac{d\sigma_{L}^{\gamma p}}{dM^2}\sim\frac{1}{M^{8}R_{0}^{4}}+\frac
{1}{m^{8}R_{0}^{4}}\ln\frac{1}{M^2R_{0}^{2}}+\left(  \frac{1}{M^2R_{0}%
^{2}}\right)  ^{3}\frac{1}{M^{6}R_{0}^{2}}+\frac{1}{M^{8}R_{0}^{4}}.
\end{equation}
Therefore%
\begin{equation}
x_{P}F_{L}^{D(3)}\sim\frac{\beta^{3}}{R_{0}^{4}}%
\end{equation}
and this contribution comes from the whole region in $x.$ In the small $\beta$
region $Q^{2}R_{0}^{2}\gg1,$ $M^2R_{0}^{2}\gg1,$ $M^2\gg Q^{2}\gg\frac
{1}{R_{0}^{2}}$ for the transverse cross section we have%
\begin{align}
&  \int rdrJ_{1}(\sqrt{x\bar{x}}Mr)K_{1}(\sqrt{x\bar{x}}Qr)\hat{\sigma}%
(r)\sim\frac{\sqrt{x\bar{x}}Mr}{\sqrt{x\bar{x}}Qr}\left(  \int_{0}^{\frac
{2}{M}}+\int_{\frac{2}{M}}^{\frac{1}{x\bar{x}M^2}}\right)  dr^{2}\hat
{\sigma}(r)\nonumber\\
&  \sim\frac{M}{Q}\left(  \int_{0}^{\left(  \frac{2}{M}\right)  ^{2}}%
\frac{r^{2}}{R_{0}^{2}}dr^{2}+\theta\left(x>\frac{1}{M^2R_{0}^{2}}\right)\int_{\left(
\frac{2}{M}\right)  ^{2}}^{\frac{1}{xM^2}}\frac{r^{2}}{R_{0}^{2}}%
dr^{2}+\theta\left(x<\frac{1}{M^2R_{0}^{2}}\right)\int_{R_{0}^{2}}^{\frac{1}{xM^2}%
}dr^{2}\right) \nonumber\\
&  \sim\frac{M}{Q}\left[  \frac{1}{M^{4}R_{0}^{2}}+\theta(x>\frac{1}%
{M^2R_{0}^{2}})\frac{1}{x^{2}M^{4}R_{0}^{2}}+\theta(x<\frac{1}{M^2%
R_{0}^{2}})\frac{1}{xM^2}\right]  ,
\end{align}
where%
\begin{equation}
J_{1}(\sqrt{x\bar{x}}Mr)\sim\sqrt{x\bar{x}}Mr\theta(1-\sqrt{x\bar{x}}Mr),\quad
K_{1}(\sqrt{x\bar{x}}Qr)\sim\frac{1}{\sqrt{x\bar{x}}Qr}.
\end{equation}
\begin{eqnarray}
&&\frac{d\sigma_{T}^{\gamma p}}{dM^2}    \sim\frac{M^2}{Q^{2}}\int_{0}%
^{1}(x\bar{x})^{2}dx\left[\left(  \frac{1}{M^{4}R_{0}^{2}}\right)  ^{2}%
+\theta(x>\frac{1}{M^2R_{0}^{2}})\frac{1}{x^{2}}\left(  \frac{1}{M^{4}%
R_{0}^{2}}\right)  ^{2}\right.\nonumber\\
&&+\theta\left(x    >\frac{1}{M^2R_{0}^{2}}\right)\left(  \frac{1}{x^{2}M^{4}R_{0}^{2}%
}\right)  ^{2}+\theta\left(x<\frac{1}{M^2R_{0}^{2}}\right)\left(  \frac{1}{xM^2%
}\right)  \frac{1}{M^{4}R_{0}^{2}}\nonumber\\
&&\left.+\theta\left(x    <\frac{1}{M^2R_{0}^{2}}\right)\left(  \frac{1}{xM^2}\right)
^{2}\right],
\end{eqnarray}%
i.e.
\begin{equation}
\frac{d\sigma_{T}^{\gamma p}}{dM^2}\sim\frac{M^2}{Q^{2}}\left[  \frac
{1}{M^{8}R_{0}^{4}}+\frac{M^2R_{0}^{2}}{M^{8}R_{0}^{4}}+\left(  \frac
{1}{M^2R_{0}^{2}}\right)  ^{2}\frac{1}{M^{6}R_{0}^{2}}+\frac{1}{M^{4}}%
\frac{1}{M^2R_{0}^{2}}\right]  .
\end{equation}%
\begin{equation}
x_{P}F_{T}^{D(3)}=\frac{Q^{4}}{4\pi^{2}\alpha\beta}\frac{d\sigma_{T,L}^{\gamma
p\rightarrow Xp^{\prime}}}{dM^2}\sim\frac{1}{\beta}\frac{1}{M^{4}R_{0}^{2}%
}\sim\frac{\beta}{R_{0}^{2}}.
\end{equation}
Again this dominant contribution comes from $x<\frac{1}{M^2R_{0}^{2}},$
i.e. aligned jets.

\section{Dipole-double dipole interference terms}
\label{append:interference}

Unfortunately \cite{Boussarie:2016ogo} does not contain expressions for
interference terms necessary for calculation of $d\sigma_{4}.$ We present them
here. The calculation is straightforward and goes along the lines described in
\cite{Boussarie:2016ogo}. In the notation of that paper the result reads
\begin{equation}
\Phi_{4}^{+}(p_{1\bot},p_{2\bot},p_{3\bot})\,\Phi_{3}^{+}(p_{1\bot}^{\prime
},p_{2\bot}^{\prime})^{\ast}=\Phi_{4}^{+}(p_{1\bot},p_{2\bot},p_{3\bot}%
)\,\Phi_{4}^{+}(p_{1\bot}^{\prime},p_{2\bot}^{\prime},0)^{\ast}+C^{++},
\end{equation}%
\begin{align}
C^{++}  &  =\frac{8p_{\gamma}^{+}{}^{4}}{z\left(  x_{q}+z\right)  \left(
\frac{\vec{p}{}_{\bar{q}2}^{\,\,2}}{x_{\bar{q}}\left(  x_{q}+z\right)  }%
+Q^{2}\right)  \left(  \frac{\vec{p}{}_{q1}^{\,\,2}}{x_{q}}+\frac{\vec{p}%
{}_{\bar{q}2}^{\,\,2}}{x_{\bar{q}}}+\frac{\vec{p_{g3}}{}^{2}}{z}+Q^{2}\right)
}\nonumber\\
&  \times\left\{  \frac{\left(  4x_{q}x_{\bar{q}}+z(2-dz)\right)  (\vec{p_{g}%
}{}-\frac{z}{x_{\bar{q}}}\vec{p}_{\bar{q}})(x_{q}\vec{p}_{g3}-z\vec{p}_{q1}%
)}{(\vec{p}_{g}-\frac{z\vec{p}_{\bar{q}}}{x_{\bar{q}}}){}^{2}\left(
\frac{\vec{p}{}_{q1^{\prime}}^{\,\,2}}{x_{q}\left(  x_{\bar{q}}+z\right)
}+Q^{2}\right)  }\right. \nonumber\\
&  -\left.  \frac{x_{\bar{q}}\left(  dz^{2}+4x_{q}\left(  x_{q}+z\right)
\right)  ({}p_{g}-\frac{z}{x_{q}}\vec{p}_{q})(\vec{p_{g3}}-\frac{z}{x_{q}}%
\vec{p}_{q1})}{(\vec{p}_{g}-\frac{z\vec{p}_{q}}{x_{q}}){}^{2}\left(
\frac{\vec{p}{}_{\bar{q}2^{\prime}}^{\,\,2}}{x_{\bar{q}}\left(  x_{q}%
+z\right)  }+Q^{2}\right)  }\right\} \nonumber\\
&  +(p_{q}\leftrightarrow p_{\bar{q}},\,p_{1}^{(\prime)}\leftrightarrow
p_{2}^{(\prime)},\,x_{q}\leftrightarrow x_{\bar{q}}).
\end{align}%
\begin{equation}
\Phi_{4}^{i}(p_{1\bot},p_{2\bot},p_{3\bot})\,\Phi_{3}^{k}(p_{1\bot}^{\prime
},p_{2\bot}^{\prime})^{\ast}=\Phi_{4}^{i}(p_{1\bot},p_{2\bot},p_{3\bot}%
)\,\Phi_{4}^{k}(p_{1\bot}^{\prime},p_{2\bot}^{\prime},0)^{\ast}+C^{ik},
\end{equation}%
\begin{align}
C^{ik}  &  =\frac{2p_{\gamma}^{+}{}^{2}}{\vec{\Delta}{}_{q}^{2}\left(
Q^{2}+\frac{\vec{p}_{g3}^{\,\,2}}{z}+\frac{\vec{p}{}_{q1}^{\,\,2}}{x_{q}%
}+\frac{\vec{p}{}_{\bar{q}2}^{\,\,2}}{x_{\bar{q}}}\right)  \left(  Q^{2}%
+\frac{\vec{p}{}_{\bar{q}2^{\prime}}^{\,\,2}}{\left(  z+x_{q}\right)
x_{\bar{q}}}\right)  }\nonumber\\
&  \times\left[  \frac{\left(  (d-2)z-2x_{q}\right)  x_{q}}{\left(
z+x_{q}\right)  {}^{3}}\left(  g_{\bot}^{ik}(\vec{p}_{\bar{q}2^{\prime}}%
\vec{\Delta}_{q})+p_{\bar{q}2^{\prime}}{}_{\bot}^{i}\Delta_{q\bot}{}%
^{k}+p_{\bar{q}2^{\prime}\bot}{}^{k}\Delta_{q\bot}{}^{i}\left(  1-2x_{\bar{q}%
}\right)  \right)  \right. \nonumber\\
&  +\frac{x_{q}\left(  \left(  (d-4)z-2x_{q}\right)  \left(  g_{\bot}%
^{ik}(\vec{p}_{\bar{q}2^{\prime}}\vec{\Delta}_{q})+p_{\bar{q}2^{\prime}\bot
}^{i}{}\Delta_{q\bot}^{k}{}\right)  +p_{\bar{q}2^{\prime}\bot}^{k}{}%
\Delta_{q\bot}^{i}{}\left(  dz+2x_{q}\right)  \left(  1-2x_{\bar{q}}\right)
\right)  }{\left(  z+x_{q}\right)  {}^{2}\left(  z+x_{\bar{q}}\right)
}\nonumber\\
&  -\frac{1}{z\left(  z+x_{q}\right)  {}^{2}x_{\bar{q}}\left(  z+x_{\bar{q}%
}\right)  {}^{2}\left(  Q^{2}+\frac{\vec{p_{q1}}{}^{2}}{x_{q}\left(
z+x_{\bar{q}}\right)  }\right)  }\left\{  z((d-4)z+2)\frac{{}}{{}}\right.
\nonumber\\
&  \times\left[  p_{q1}{}_{\bot}^{i}\left(  (\vec{p}_{\bar{q}2^{\prime}}%
\vec{\Delta}_{q})P_{\bot}^{k}-(\vec{P}\vec{p}_{\bar{q}2^{\prime}})\Delta_{q}%
{}_{\bot}^{k}\right)  \left(  2x_{q}-1\right)  -(\vec{P}\vec{p}_{\bar
{q}2^{\prime}})\left(  g_{\bot}^{ik}(\vec{p}_{q1}\vec{\Delta}_{q})+p_{q1}%
{}_{\bot}^{k}\Delta_{q}{}_{\bot}^{i}\right)  \right. \nonumber\\
&  -\!\left.  P_{\bot}^{k}\!\left(  \!(\vec{p}_{q1}\vec{\Delta}_{q})p_{\bar
{q}2^{\prime}}{}_{\bot}^{i}\!-\!(\vec{p}_{q1}\vec{p}_{\bar{q}2^{\prime}%
})\Delta_{q}{}_{\bot}^{i}\!\right)  \!\right]  +4x_{q}z\left(  1-2x_{q}%
\right)  p_{q1}{}_{\bot}^{i}\!\left(  \!(\vec{p}_{\bar{q}2^{\prime}}%
\vec{\Delta}_{q})P_{\bot}^{k}-(\vec{P}\vec{p}_{\bar{q}2^{\prime}})\Delta_{q}%
{}_{\bot}^{k}\!\right) \nonumber\\
&  +z\left(  1-2x_{\bar{q}}\right)  \left(  dz+4x_{q}-2\right)  p_{\bar
{q}2^{\prime}}{}_{\bot}^{k}\left(  (\vec{p}_{q1}\vec{\Delta}_{q})P_{\bot}%
^{i}-(\vec{P}\vec{p}_{q1})\Delta_{q}{}_{\bot}^{i}\right)
-z((d-4)z-2)\nonumber\\
&  \times\left[  \left(  g_{\bot}^{ik}(\vec{P}\vec{p}_{q1})+P_{\bot}^{i}%
p_{q1}{}_{\bot}^{k}\right)  (\vec{p}_{\bar{q}2^{\prime}}\vec{\Delta}%
_{q})+\left(  (\vec{P}\vec{p}_{q1})p_{\bar{q}2^{\prime}}{}_{\bot}^{i}-(\vec
{p}_{q1}\vec{p}_{\bar{q}2^{\prime}})P_{\bot}^{i}\right)  \Delta_{q}{}_{\bot
}^{k}\right] \nonumber\\
&  +(\vec{P}\vec{\Delta}_{q})p_{q1}{}_{\bot}^{i}p_{\bar{q}2^{\prime}}{}_{\bot
}^{k}\left(  1-2x_{q}\right)  \left(  1-2x_{\bar{q}}\right)  \left(
z(dz-2)-4x_{q}x_{\bar{q}}\right) \nonumber\\
&  -\left.  (\vec{P}\vec{\Delta}_{q})\left(  g_{\bot}^{ik}(\vec{p}_{q1}\vec
{p}_{\bar{q}2^{\prime}})+p_{q1}{}_{\bot}^{k}p_{\bar{q}2^{\prime}}{}_{\bot}%
^{i}\right)  \left(  z(2-(d-4)z)+4x_{q}x_{\bar{q}}\right)  \right\} \nonumber
\end{align}%
\begin{align}
&  -\frac{1}{z\left(  z+x_{q}\right)  {}^{4}\left(  Q^{2}+\frac{\vec{p}%
{}_{\bar{q}2}^{\,\,2}}{\left(  z+x_{q}\right)  x_{\bar{q}}}\right)  x_{\bar
{q}}}\left\{  z\left(  dz+4x_{\bar{q}}-4\right)  \left[  \left(  1-2x_{\bar
{q}}\right)  \frac{{}}{{}}\right.  \right. \nonumber\\
&  \times\left(  p_{\bar{q}2^{\prime}}{}_{\bot}^{k}\left(  (\vec{p}_{\bar{q}%
2}\vec{\Delta}_{q})W_{\bot}^{i}-(\vec{W}\vec{p}_{\bar{q}2})\Delta_{q}{}_{\bot
}^{i}\right)  +p_{\bar{q}2}{}_{\bot}^{i}\left(  (\vec{W}\vec{p}_{\bar
{q}2^{\prime}})\Delta_{q}{}_{\bot}^{k}-(\vec{p}_{\bar{q}2^{\prime}}\vec
{\Delta}_{q})W_{\bot}^{k}\right)  \right) \nonumber\\
&  +W_{\bot}^{k}\left(  (\vec{p}_{\bar{q}2}\vec{\Delta}_{q})p_{\bar
{q}2^{\prime}}{}_{\bot}^{i}-(\vec{p}_{\bar{q}2}\vec{p}_{\bar{q}2^{\prime}%
})\Delta_{q}{}_{\bot}^{i}\right)  +\left(  (\vec{p}_{\bar{q}2}\vec{p}_{\bar
{q}2^{\prime}})W_{\bot}^{i}-(\vec{W}\vec{p}_{\bar{q}2})p_{\bar{q}2^{\prime}}%
{}_{\bot}^{i}\right)  \Delta_{q}{}_{\bot}^{k}\nonumber\\
&  +\left.  g_{\bot}^{ik}\left(  (\vec{W}\vec{p}_{\bar{q}2^{\prime}})(\vec
{p}_{\bar{q}2}\vec{\Delta}_{q})-(\vec{W}\vec{p}_{\bar{q}2})(\vec{p}_{\bar
{q}2^{\prime}}\vec{\Delta}_{q})\right)  +p_{\bar{q}2}{}_{\bot}^{k}\left(
(\vec{W}\vec{p}_{\bar{q}2^{\prime}})\Delta_{q}{}_{\bot}^{i}-(\vec{p}_{\bar
{q}2^{\prime}}\vec{\Delta}_{q})W_{\bot}^{i}\right)  \right] \nonumber\\
&  +\left.  \left.  (\vec{W}\vec{\Delta}_{q})\left(  p_{\bar{q}2}{}_{\bot}%
^{i}p_{\bar{q}2^{\prime}}{}_{\bot}^{k}\left(  1-2x_{\bar{q}}\right)  {}%
^{2}-g_{\bot}^{ik}(\vec{p}_{\bar{q}2}\vec{p}_{\bar{q}2^{\prime}})-p_{\bar{q}%
2}{}_{\bot}^{k}p_{\bar{q}2^{\prime}}{}_{\bot}^{i}\right)  \left(
dz^{2}-4x_{q}\left(  x_{\bar{q}}-1\right)  \right)  \right\}  \frac{{}}{{}%
}\right] \nonumber\\
&  +(p_{q}\leftrightarrow p_{\bar{q}},\,p_{1}^{(\prime)}\leftrightarrow
p_{2}^{(\prime)},\,x_{q}\leftrightarrow x_{\bar{q}}),
\end{align}
where%
\begin{equation}
W_{\bot}^{i}=x_{q}p_{g3\bot}^{i}-zp_{q1\bot}^{i},\quad P_{\bot}^{i}=x_{\bar
{q}}p_{g3\bot}^{i}-zp_{\bar{q}2\bot}^{i}.
\end{equation}

\section{Normalization}

\label{append:normalization}

In this appendix we discuss the overall normalization
of the cross section and the relation of our matrix elements $F$ defined in
(5.2--8) of ref.~\cite{Boussarie:2016ogo} to the GBW dipole cross section.

The density matrix for the LO cross section in our frame (5.21-23) was
obtained in ref.~\cite{Boussarie:2016ogo}. To get the proper normalization we have
to multiply all cross sections in ref.~\cite{Boussarie:2016ogo} by $\frac{1}%
{2(2\pi)^{4}}.$ Indeed, the factor $\frac{1}{2}$ comes from the normalization
of $A_{3}$ in eq.~(5.11) of ref.~\cite{Boussarie:2016ogo}. The in and out proton states
are normalized there to have
\begin{equation}
\frac{1}{\sqrt{2p_{0}^{-}}\sqrt{2p_{0}^{\prime-}}}\simeq\frac{1}{\sqrt{4E_{0}%
}\sqrt{4E_{0}^{\prime}}}=\frac{1}{2}\frac{1}{\sqrt{2E_{0}}\sqrt{2E_{0}%
^{\prime}}}.
\end{equation}
Since the S-matrix does not depend on state normalization, $A_{3}$ is two
times bigger than the standard amplitude normalized to $\frac{1}{\sqrt{2E_{0}%
}\sqrt{2E_{0}^{\prime}}}.$ As a result, the cross section should have an extra
$\frac{1}{4}$ to compensate for it, i.e. in (5.1) of ref.~\cite{Boussarie:2016ogo} we should have had
\begin{equation}
d\sigma=\frac{1}{4}\frac{1}{2s}(2\pi)^{4}\delta^{(4)}(p_{\gamma}+p_{0}%
-p_{q}-p_{\bar{q}}-p_{0}^{\prime})\left\vert A_{3}\right\vert ^{2}d\rho_{3}.
\end{equation}
The same correction must be done in eq.~(6.1) of ref.~\cite{Boussarie:2016ogo}.

The $2\pi$ power must be corrected in eq.~(5.11) of ref.~\cite{Boussarie:2016ogo} in the overall factor%
\begin{equation}
\frac{1}{(2\pi)^{D-4}}\rightarrow\frac{1}{(2\pi)^{D-2}}.
\end{equation}
Indeed, the amplitude $A_{3}$\ is exactly the matrix element (3.1) of ref.~\cite{Boussarie:2016ogo} after removing
$(2\pi)^{4}\delta^{(4)}(p_{\gamma}+p_{0}-p_{q}-p_{\bar{q}}-p_{0}^{\prime}).$
In this matrix element transverse and $^{(-)}$ delta functions appear together
with $(2\pi)^{2}$ and $2\pi$ as eqs.~(5.7--8) and eqs.~(5.2--3) of ref.~\cite{Boussarie:2016ogo} correspondingly. Only
the $^{(+)}$ delta function is without $2\pi$ in eq.~(3.1). Therefore we must have
an extra $2\pi$ in the denominator in $A_{3}$ in addition to $\frac{1}%
{(2\pi)^{D-3}}$ from eq.~(3.1) of ref.~\cite{Boussarie:2016ogo}. This gives us the aforementioned substitution. The
same misprint was done in eq.~(6.4) of ref.~\cite{Boussarie:2016ogo}. After these corrections we get
eqs.~(\ref{sigmatt0}--\ref{sigmatt}).

Next, we have to substitute a model for the hadronic matrix elements
$\mathbf{F}.$ We will use the Golec-Biernat - W\"usthoff (GBW)
\cite{GolecBiernat:1998js}\ parametrization, which was formulated in the
coordinate space. To get the proper normalization we Fourier transform eq.~(\ref{sigmall}) and compare it with Eq.~(4.48) in
ref.~\cite{GolecBiernat:2001hab}. Using%
\begin{equation}
\frac{1}{\vec{l}^{2}+a^{2}}=\int d^{2}r\frac{K_{0}(ar)}{2\pi}e^{-i\vec{l}%
\vec{r}},\quad\mathbf{F}(\vec{k})=\int d\vec{r}e^{-i\vec{k}\vec{r}}F(\vec{r}),
\end{equation}
we have%
\begin{equation}
\left.  \frac{d\sigma_{0LL}}{dxd\vec{p}_{q}d\vec{p}_{\bar{q}}}\right\vert
_{t=0}=\frac{1}{2(2\pi)^{4}}\frac{4\alpha Q_{q}^{2}}{N_{c}}x^{2}\bar{x}%
^{2}Q^{2}\left\vert \int d^{2}r\frac{K_{0}(\sqrt{x\bar{x}}Qr)}{2\pi}%
e^{i\frac{p_{q\bar{q}}{}_{\bot}}{2}\vec{r}}F(\vec{r})\right\vert ^{2}%
\end{equation}
and
\begin{equation}
\left.  \frac{d\sigma_{0LL}}{dt}\right\vert _{t=0}=\frac{1}{2(2\pi)^{4}}%
\frac{4\alpha Q_{q}^{2}}{N_{c}}\pi\int dxQ^{2}x^{2}\bar{x}^{2}\int d^{2}%
rK_{0}(\sqrt{x\bar{x}}Qr)^{2}F(\vec{r})^{2}.
\end{equation}
Comparing it with eq.~(4.48) in ref.~\cite{GolecBiernat:2001hab}, the GBW
parametrization of the forward dipole matrix element in our normalization
reads%
\begin{align}
F_{p_{0\bot}p_{0\bot}}(z_{\bot}) &  =\left.  \frac{\langle P^{\prime}%
(p_{0}^{\prime})|T(\Tr(U_{\frac{z_{\bot}}{2}}U_{-\frac{z_{\bot}}{2}}^{\dag
})-N_{c})|P(p_{0})\rangle}{2\pi\delta(p_{00^{\prime}}^{-})}\right\vert
_{p_{0}\rightarrow p_{0}^{\prime}}\nonumber\\
&  =F(z_{\bot})=N_{c}\sigma_{0}(1-e^{-\frac{z^{2}}{4R_{0}^{2}}}).\label{F}%
\end{align}
One can check the consistency of this normalization 
by deriving the inclusive $\gamma^{\ast}p$ cross section with the same matrix elements. Using propagators in the shockwave background
(2.19--20) from ref.~\cite{Boussarie:2016ogo}, one gets for the $\gamma^{\ast
}p\rightarrow\gamma^{\ast}p$ amplitude%
\begin{equation}
iA=\sqrt{2p_{0}^{-}2p_{0}^{\prime-}}Q_{q}^{2}(-ie)^{2}\int dx\int
dy \Tr[\hat{\varepsilon}_{1}G(x-y)\hat{\varepsilon}_{2}G(y-x)]e^{-ip_{\gamma
}x+ip_{\gamma}^{\prime}y}%
\end{equation}%
\begin{align}
&  =\frac{8\alpha Q_{q}^{2}p_{\gamma}^{+}}{\pi}\delta(p_{\gamma}^{\prime
+}-p_{\gamma}^{+})\int dxQ^{2}x^{2}\bar{x}^{2}K_{0}(r_{12}\sqrt{x\bar
{x}Q_{\gamma}^{2}})K_{0}(r_{12}\sqrt{x\bar{x}Q_{\gamma^{\prime}}^{2}%
})\nonumber\\
&  \times\sqrt{2p_{0}^{-}2p_{0}^{\prime-}}\int d^{D-2}r_{2\bot}\int
d^{D-2}r_{1\bot}\langle P^{\prime}(p_{0}^{\prime})|T(\Tr[U_{1}U_{2}^{\dag
}]-N_{c})|P(p_{0})\rangle.
\end{align}
Extracting the dependence on the overall momentum transfer
\begin{align}
\langle P^{\prime}(p_{0}^{\prime})|T \Tr[U_{1}U_{2}^{\dag}]|P(p_{0})\rangle &
=\langle p_{0}^{\prime}|e^{\pm i\hat{P}\frac{(r_{1}+r_{2})}{2}}T \,\Tr[U_{r_{1}%
}U_{r_{2}}^{\dag}]e^{\mp i\hat{P}\frac{(r_{1}+r_{2})}{2}}|p_{0}\rangle
\nonumber\\
&  =e^{ip_{0^{\prime}0\bot}\frac{(r_{1}+r_{2})_{\bot}}{2}}\langle
p_{0}^{\prime}|T \, \Tr[U_{\frac{r_{12\bot}}{2}}U_{-\frac{r_{12\bot}}{2}}^{\dag
}]|p_{0}\rangle,
\end{align}
we get%
\begin{align}
iA &  =(2\pi)^{4}\delta(p_{\gamma}^{\prime+}-p_{\gamma}^{+})\delta
(p_{0^{\prime}0\bot})\delta(p_{00^{\prime}}^{-})\int d^{2}z\frac{\langle
P^{\prime}(p_{0}^{\prime})|T(\Tr[U_{\frac{z}{2}}U_{-\frac{z}{2}}^{\dag}%
]-N_{c})|P(p_{0})\rangle}{2\pi\delta(p_{00^{\prime}}^{-})}\nonumber\\
&  \times\sqrt{2p_{0}^{-}2p_{0}^{\prime-}}\frac{4\alpha Q_{q}^{2}p_{\gamma
}^{+}}{\pi^{2}}\int dxQ^{2}x^{2}\bar{x}^{2}K_{0}(z\sqrt{x\bar{x}Q_{\gamma}%
^{2}})K_{0}(z\sqrt{x\bar{x}Q_{\gamma^{\prime}}^{2}}).
\end{align}
Then, using the optical theorem%
\begin{align}
\sigma_{tot} &  =\frac{\operatorname{Im}A_{ii}}{2s}=\int d^{2}z\left.
\frac{\langle P^{\prime}(p_{0}^{\prime})|T(\Tr[U_{\frac{z}{2}}U_{-\frac{z}{2}%
}^{\dag}]-N_{c})|P(p_{0})\rangle}{2\pi\delta(p_{00^{\prime}}^{-})}\right\vert
_{p_{0}^{\prime}\rightarrow p_{0}}\nonumber\\
&  \times\frac{\alpha Q_{q}^{2}}{2\pi^{2}}\int dx4Q^{2}x^{2}\bar{x}^{2}%
K_{0}^{2}(z\sqrt{x\bar{x}Q_{\gamma}^{2}}).
\end{align}
Comparing this result to eqs.~(3.7--9) in ref.~\cite{GolecBiernat:2001hab}, we get the
same result (\ref{F}) for F as before.

\providecommand{\href}[2]{#2}\begingroup\raggedright\endgroup

\end{document}